\documentclass[11pt]{article}
\usepackage{jheppub}

\usepackage{amsmath,amsfonts,amsbsy,amssymb,array,accents,dsfont}
\usepackage{enumerate,array,latexsym,graphicx,mathrsfs,verbatim,psfrag}
\usepackage{enumerate}
\usepackage{graphicx} 

\usepackage{bm}
\usepackage{amssymb}
\usepackage{amsmath}
\usepackage{cancel}
\usepackage{epigraph}

\newenvironment{centermath}
 {\begin{center}$\displaystyle}
 {$\end{center}}

\makeatletter
\DeclareFontFamily{OMX}{MnSymbolE}{}
\DeclareSymbolFont{MnLargeSymbols}{OMX}{MnSymbolE}{m}{n}
\SetSymbolFont{MnLargeSymbols}{bold}{OMX}{MnSymbolE}{b}{n}
\DeclareFontShape{OMX}{MnSymbolE}{m}{n}{
    <-6>  MnSymbolE5
   <6-7>  MnSymbolE6
   <7-8>  MnSymbolE7
   <8-9>  MnSymbolE8
   <9-10> MnSymbolE9
  <10-12> MnSymbolE10
  <12->   MnSymbolE12
}{}
\DeclareFontShape{OMX}{MnSymbolE}{b}{n}{
    <-6>  MnSymbolE-Bold5
   <6-7>  MnSymbolE-Bold6
   <7-8>  MnSymbolE-Bold7
   <8-9>  MnSymbolE-Bold8
   <9-10> MnSymbolE-Bold9
  <10-12> MnSymbolE-Bold10
  <12->   MnSymbolE-Bold12
}{}

\let\llangle\@undefined
\let\rrangle\@undefined
\DeclareMathDelimiter{\llangle}{\mathopen}%
                     {MnLargeSymbols}{'164}{MnLargeSymbols}{'164}
\DeclareMathDelimiter{\rrangle}{\mathclose}%
                     {MnLargeSymbols}{'171}{MnLargeSymbols}{'171}
\makeatother

\def\be{ \begin{equation} }
\def\ee{ \end{equation}}

\newcommand{\eq}[1]{\begin{align}\begin{split}#1\end{split}\end{align}}

\def\cot{{\rm cot}}
\def\det{{\rm det}}

\def\exp{{\rm exp}}

\def\half{\frac{1}{2}}

\def\one{{\hbox{ 1\kern-.8mm l}}}
\def\CA{\cal A}

\def\CS {{\cal S}}

\def\IN{\mathbb{N}}

\def\IR{{\mathbb{R}}}
\def\IS{{\mathbb{S}}}
\def\IT{{\mathbb{T}}}

\def\IZ{{\mathbb{Z}}}

\def\ff{\mathfrak{f}}

\def\rmk#1{\bigskip\noindent{\bf Remark} }
\def\cnj#1{\bigskip\noindent{\bf Conjecture:} }
\DeclareMathAlphabet{\mathpzc}{OT1}{pzc}{m}{it}

\DeclareMathSymbol{\medhatsym}{\mathord}{largesymbols}{"62} % basic symbol
\newcommand\lowermedhatsym{%  adjust height
  \text{\smash{\raisebox{-1.28ex}{%
    $\medhatsym$}}}}
\newcommand\medhat[1]{% command to be used
  \mathchoice
    {\accentset{\displaystyle\lowermedhatsym}{#1}}
    {\accentset{\textstyle\lowermedhatsym}{#1}}
    {\accentset{\scriptstyle\lowermedhatsym}{#1}}
    {\accentset{\scriptscriptstyle\lowermedhatsym}{#1}}
}

\DeclareMathSymbol{\medtildesym}{\mathord}{largesymbols}{"65}% basic symbol

\makeatletter
\renewcommand\section{\@startsection {section}{1}{\z@}%
                                   {-3.5ex \@plus -1ex \@minus -.2ex}%nn
                                   {2.3ex \@plus.2ex}%
                                   {\normalfont\Large\bfseries}}
\renewcommand\subsection{\@startsection{subsection}{2}{\z@}%
                                     {-3.25ex\@plus -1ex \@minus -.2ex}%
                                     {1.5ex \@plus .2ex}%
                                     {\normalfont\bfseries}}
\renewcommand\subsubsection{\@startsection{subsubsection}{3}{\z@}%
                                     {-2.5ex\@plus -1ex \@minus -.2ex}%
                                     {1.25ex \@plus .2ex}%
                                     {\normalfont\textit}}
\makeatother
\def\tight#1{\! #1 \!}

\def\({\left(}
\def\){\right)}
\def\[{\left[}
\def\]{\right]}

\def\naive{na\"ive}
\def\sl{\text{sl}}
\def\su{\text{su}}

\def \sutil{{\widetilde{\text{su}}}}
\def\sltwo{{SL(2,\IR)}}

\def\sutwo{{SU(2)}}
\def\sutwotil{{\widetilde{SU}(2)}}
\def\uone{U(1)}

\def\coset{{\cal{K}}}
\def\matter{{\cal{M}}}
\def\Neqtwo{\cN\tight=2}
\def\dlam{{\delta\hskip -1pt\lambda}}

\def\ie{{i.e.}}
\def\eg{{e.g.}}

\def\lstr{\ell_{\textit{str}}}

\def\gstrsq{g_{\textit s}^{2}}

\def\rads{R_{\textit AdS}}

\def\nfive{{n_5}}

\def\nfivetil{{\tilde n_5}}
\def\nfivehat{{\hat n_5}}

\def\nz{k_\sl}
\def\nf{k_\su}
\def\nft{k_\sutil}
\def\ksl{k_\sl}
\def\hatksl{{\hat k_\sl}}
\def\ksu{k_\su}
\def\tilksu{{\tilde k_\su}}

\def\sst#1{\scriptscriptstyle{#1}}

\def\J{{\mathsf J}}
\def\M{{\mathsf M}}

\def\sfX{{\mathsf X}}

\def\sfZ{{\mathsf Z}}
\def\sfv{{\mathsf v}}

\def\x1x2{$x^1$-$x^2$}

\newcommand{\cAb}{\ensuremath{\bar{\mathcal{A}}}}

\def\gtil{{\tilde g}}
\def\ytil{{\tilde y}}

\def\Ry{R_y}
\def\Rytil{R_\ytil}

\def\alphab{{\boldsymbol\alpha}}

\def\lamb{{\boldsymbol\lambda}}

\def\phil{{\varphi_l}}
\def\phir{{\varphi_r}}
\def\philt{{\tilde\varphi_l}}
\def\phirt{{\tilde\varphi_r}}
\def\phit{\tilde\phi}
\def\psit{\tilde\psi}
\def\thetat{{\tilde\theta}}
\def\chihat{{\hat\chi}}

\def\mub{{\boldsymbol\mu}}

\def\Gtil{{\widetilde G}}
\def\Btil{{\widetilde B}}
\def\vareps{{\varepsilon}}

\def\Psihat{{\medhat\Psi}}
\def\Phihat{{\medhat\Phi}}
\def\jsu{j_\su}
\def\jsut{j_{\sutil}}
\def\msu{m_\su}
\def\msut{m_{\sutil}}

\def\bmsu{\bar m_\su}
\def\bmsut{\bar m_{\sutil}}

\def\jsl{j_\sl}
\def\msl{m_\sl}
\def\wsl{w_\sl}
\def\bmsl{\bar m_\sl}
\def\bwsl{\bar w_\sl}
\def\Jsq{J^\flat}
\def\Jbsq{{\bar J}^\flat}
\def\ff{f}
\def\fft{\tilde f}

\def\rhotil{\tilde\rho}

\def\sst{\scriptscriptstyle}
\def\half{\frac12}
\def\coeff#1#2{{\textstyle \frac{#1}{#2}}}
\def\hf{\coeff12}

\def\One{{\hbox{1\kern-1mm l}}}

\def\barray{\begin{array}}
\def\earray{\end{array}}
\def\be{\begin{equation}}
\def\ee{\end{equation}}
\def\bea{\begin{eqnarray}}
\def\eea{\end{eqnarray}}
\def\bal{\begin{align}}
\def\eal{\end{align}}
\def\nn{\nonumber}

\newcommand{\bC}{{\mathbb C}}

\newcommand{\bR}{{\mathbb R}}
\newcommand{\bS}{{\mathbb S}}
\newcommand{\bT}{{\mathbb T}}
\newcommand{\bZ}{{\mathbb Z}}

\def\IN{\mathbb{N}}

\def\IR{\mathbb{R}}
\def\IS{\mathbb{S}}
\def\IT{\mathbb{T}}
\def\IZ{\mathbb{Z}}

\numberwithin{equation}{section}

\mathchardef\mhyphen="2D

\def\cA{\mathcal {A}}  \def\cC{\mathcal {C}}
\def\cD{\mathcal {D}} \def\cE{\mathcal {E}} 
\def\cG{\mathcal {G}} \def\cH{\mathcal {H}} 
\def\cJ{\mathcal {J}} \def\cK{\mathcal {K}} \def\cL{\mathcal {L}}
\def\cM{\mathcal {M}} \def\cN{\mathcal {N}} 
 \def\cQ{\mathcal {Q}} \def\cR{\mathcal {R}}
\def\cS{\mathcal {S}}  
\def\cV{\mathcal {V}} \def\cW{\mathcal {W}} 
\def\cY{\mathcal {Y}}

\def\one{{\hbox{\kern+.5mm 1\kern-.8mm l}}}
\def\zero{{\hbox{0\kern-1.5mm 0}}}

\newcommand{\ket}[1]{{\,| {#1} \rangle}}

\newcommand{\T}[3]{\ensuremath{ #1{}^{#2}_{\phantom{#2} \! #3}}}		%general tensor with upper indices displayed first

\title{Wrapped Fivebranes Redux}

\author{T.~Daniel Brennan and Emil J. Martinec}

\affiliation{Kadanoff Center for Theoretical Physics \& Enrico Fermi Institute, \\
University of Chicago,
Michelson Center for Physics, 933 E 56th St, Chicago, IL 60637. 
}

\emailAdd{tdbrennan@uchicago.edu, e-martinec@uchicago.edu}

\abstract{The near-horizon region of Neveu-Schwarz fivebranes provides interesting examples of gauge/gravity duality.
We revisit the structure of wrapped and/or intersecting fivebranes using the tools of null-gauged WZW models in worldsheet string theory, revealing the effective geometry of the fivebrane throat in a variety of examples.  Variant gaugings yield linear dilaton fivebrane throats with $AdS_3$ caps, providing a wealth of information about the near-BPS structure of the corresponding spacetime CFT duals.
}

\begin{document}

\maketitle

%%%%%%%%%%%%%%%%%%%%%%%%%%%%%%%%%%%%%%%%%%%%%%%%
%%%%%%%%%%%%%%%%%%%%%%%%%%%%%%%%%%%%%%%%%%%%%%%%

\section{Introduction}
\label{sec:intro}

Gauged Wess-Zumino-Witten (WZW) models have proven to be a valuable tool for studying string theory beyond the supergravity approximation (for a review, see for example~\cite{Tseytlin:1995fh}).  The underlying current algebra structure allows the string worldsheet dynamics to be solved exactly~\cite{Knizhnik:1984nr,Gepner:1987qi,Kazama:1988qp}. 
These so-called coset models can be used to describe stringy regimes of the moduli space of string compactifications through the so-called Calabi-Yau/Landau-Ginsburg correspondence~\cite{Gepner:1987qi,Martinec:1988zu,Vafa:1988uu,Greene:1988ut}.  Noncompact versions of this correspondence describe the throats of Neveu-Schwarz (NS) fivebranes~\cite{Callan:1991at,Sfetsos:1998xd,Giveon:1999px}, as well as the Euclidean black NS5-brane~\cite{Kutasov:2000jp}.  Lorentzian cosets have been proposed to describe 2d dilatonic black holes~\cite{Witten:1991yr,Dijkgraaf:1991ba}, and model cosmologies~\cite{Nappi:1992kv,Kounnas:1992wc,Kiritsis:1994np,Elitzur:2002rt}, as well as the Lorentzian black fivebrane (although here one may question whether string perturbation theory is under control due to the presence of singularities and/or closed timelike curves).

Recently, the gauging of null isometries has been used to tie together some of these constructions, as well as generate new ones~-- for instance the construction of solvable worldsheet descriptions of families of heavy states in $AdS_3$~\cite{Martinec:2017ztd,Martinec:2018nco,Martinec:2019wzw,Martinec:2020gkv}.
The common theme of these constructions is that the Wess-Zumino term of the sigma model on a group manifold is a natural source of electric and magnetic Neveu-Schwarz (NS) three-form flux.  For instance, a single stack of fivebranes is described by the quotient
\be
\label{basic coset}
\frac{\cG}{\cH} = \frac{\sltwo\times\sutwo\times \bR^{d-1,1}}{\uone_\ell\times\uone_r}
\ee
where the left and right $\uone$ group actions being gauged are generated by null Killing vectors.  If the isometries act entirely within $\sltwo\times\sutwo$, then the $H$ flux in the effective geometry is magnetic, and one has just fivebranes; if they act as well on the $\bR^{d-1,1}$ then one can arrange that the background carries electric $H$ flux as well.  These quotients were considered in the critical dimension (where $d=6$) in~\cite{Martinec:2017ztd,Martinec:2018nco,Martinec:2019wzw,Martinec:2020gkv}; we consider the non-critical versions for $d=4$ and $d=2$ in section~\ref{sec:ArgyresDouglas}.  These lower-dimensional versions are thought to describe NS5-branes wrapped on the vanishing cycle of a local Calabi-Yau singularity~\cite{Giveon:1999zm,Giveon:1999px,Giveon:1999tq}
\be
\sfZ_1^{k_\su} + \sfZ_2^{\,2} + \dots + \sfZ_{\frac{12-d}2}^{\,2} = \lambda  
\ee
where $\lambda$ parametrizes a tiny resolution of the singularity;
the idea is that the vanishing cycle wrapped by the fivebrane is so sub-stringy in size that there is no room to wiggle the string along the cycle, and so the worldsheet field theory sees an effectively reduced dimensionality of the target space.

Another theme of this construction is the infrared regularization of string dynamics in the vicinity of NS5-brane sources.  The throat of coincident fivebranes has a strong coupling singularity which precludes a consistent perturbative string S-matrix.  But gauged WZW models such as~\eqref{basic coset} are perfectly well-behaved -- the WZW model on $\cG$ is completely non-singular, and the gauging of $\cH$ merely selects the subset of amplitudes that involve $\cH$-invariant observables.  The reason the model is non-singular is that the coset models don't describe coincident fivebranes, but rather fivebranes that have been separated slightly on their Coulomb branch.  There is no perturbative string dynamics in the throat of a single fivebrane~-- in the standard covariant worldsheet formalism, the supersymmetric $\sutwo$ WZW model that describes the angular degrees of freedom of an NS5 throat has a minimum of two quanta of $H$ flux.%
\footnote{There is a recent proposal~\cite{Eberhardt:2018ouy} to describe a fivebrane throat with a single unit of $H$ flux using a hybrid Green-Schwarz formalism~\cite{Berkovits:1999im}, but it seems that the string wavefunctions are pinned near the asymptotic boundary, and there are no operators that create strings propagating down the throat.  So again while there is nominally a description of the fivebrane throat, there is no radial dynamics, and the string wavefunctions have no support near the fivebrane.}
When a string is sent down the throat toward the fivebrane source, it gets to the radial scale set by the fivebrane separation and it can't get closer to any of the separated branes because of this fact; instead the string bounces off of an effective wall that shields it from the strong coupling region~\cite{Giveon:1999px,Giveon:1999tq}.  We will see in this class of examples that in the effective geometry, the fivebranes are indeed always separated out into a circular array in their transverse space.

A second class of examples, considered in sections~\ref{sec:1d-2d intersection} and~\ref{sec:3d intersection}, involves fivebrane intersections~\cite{Itzhaki:2005tu,Elitzur:1998mm,Dasgupta:1998su,PandoZayas:2000he}.  In this case, a cycle in the angular space of one stack of fivebranes is wrapped by the other stack of fivebranes.  We will again see that the intersection loci of the fivebranes are separated and spread out along a circular array.  Of particular interest is the situation discussed in section~\ref{sec:3d intersection}, where the fivebranes intersect in codimension four, over $\bR^{3,1}$.  This situation generates interesting 4d gauge theories on the intersection, and some of their aspects should be in reach of worldsheet techniques.  The class of gauged WZW models involves
\be
\label{two stack coset}
\frac{\cG}{\cH} = \frac{\sltwo\times\sutwo\times \sutwotil \times \bR^{3,1}}{\uone_\ell\times\uone_r\times \uone_s}
\ee
where $\uone_s$ is an additional spatial isometry.  This class of models includes the examples described in~\cite{Giveon:1999px} involving fivebranes wrapping more complicated hypersurface singularities 
\be
\sfZ_1^{k_\su} + \sfZ_2^{k_\sutil} + \sfZ_3^{\,2} + \dots + \sfZ_{\frac{12-d}2}^{\,2} = \lambda  ~.
\ee
In the null gauging presentation~\eqref{two stack coset} of this background, the spatial isometry $\uone_s$ has fixed points, and the effective sigma model target space turns out not to be a conventional geometry, rather it is a {\it T-fold}~-- a space for which, in order to maintain the appropriate low-energy effective field theory everywhere, the transitions between local patches of the target space involve T-duality transformations in addition to coordinate maps.  To arrive at something less exotic, we consider an alternative spatial isometry without fixed points~\cite{Guadagnini:1987ty,PandoZayas:2000he} which describes the throat of intersecting fivebranes and KK monopoles, a setup that has been used in brane constructions of 4d gauge theory (see~\cite{Giveon:1998sr} for a review), and related by dualities to a system of D3-branes at the tip of a conifold singularity~\cite{Dasgupta:1998su}.

In section~\ref{sec:supertubes}, we further generalize the null isometries to include contributions from the flat directions along the fivebranes, compactifying the spatial directions along a torus and tilting the null isometry partly along one of the torus cycles to make a {\it supertube} following the same route that produces two-charge supertubes out of NS5-branes wrapped on $\bT^5$~\cite{Martinec:2017ztd}.  Choosing the second charge to be that of fundamental strings wrapping $\bS^1_y\subset \bT^5$, the result is a set of backgrounds describing holographic RG flows between the linear dilaton throat of fivebranes in the UV, transitioning to an $AdS_3$ cap in the IR.  In this IR limit, the states being described are a family of 1/2-BPS states in the dual spacetime CFT; here we are getting insight into their gravitational dual description at the fully stringy level.  In addition, as discussed recently in~\cite{Martinec:2020gkv} for fivebranes wrapped on $\bS^1_y\times \bT^4$, one can identify among the string vertex operators the 1/2-BPS spectrum that takes one to nearby BPS states of the spacetime CFT.  One expects that a similarly complete picture of the near-BPS state space can be developed for the models described here.

We conclude with a discussion of further directions in section~\ref{sec:discussion}.  Several appendices lay out our conventions, the structure of spacetime supersymmetry, and an alternative non-null choice of gauge isometries that nevertheless leads to the same physical spectrum as the null gauging we employ in the body of the paper.

%%%%%%%%%%%%%%%%%%%%%%%%%%%%%%%%%%%%%%
%%%%%%%%%%%%%%%%%%%%%%%%%%%%%%%%%%%%%%

\section{Fivebranes wrapping a resolved Calabi-Yau singularity}
\label{sec:ArgyresDouglas}

We begin with a class of examples related to 4d gauge theories, namely a decoupling limit of NS5-branes wrapping a Seiberg-Witten curve of a slightly resolved conifold, whose low-energy dynamics is $\cN=2$ gauge theory near an Argyres-Douglas point~\cite{Giveon:1999zm}.  Here the coset is
\be
\label{ADcoset}
\bigg(\frac{\sltwo\times\sutwo}{\uone_\ell\times\uone_r}\bigg)  \times \bR^{3,1} ~.
\ee
One has a choice of ADE modular invariant for $\sutwo$, leading to an ADE classification of backgrounds.
The requirement that this CFT have central charge $c=15$ is satisfied if the levels $\nz$ of the $\sltwo$ factor and $\nf$ of the $\sutwo$ factor are related by
\be
\label{sltwo level}
\nz=\frac{2\nf}{\nf+2} ~.
\ee
Note that the curvature radius of $\sltwo$, which in the critical dimension is $\nfive\lstr$, is here between one and two in units of $\lstr$, and so the fivebrane throat is always quite stringy.  

%%%%%%%%%%%%%%%%%%%%%%%%%%%%%%%%%%%%%%
\subsection{Preliminaries}
\label{sec:ADsetup}

The differing levels of the current algebras leads to new wrinkles in the null gauging formalism.  The natural null currents
\be
\cJ = J^3_\sl - \sqrt{\frac{\nz}{\nf}}\, J^3_\su
~~,~~~~
\bar \cJ = \bar J^3_\sl - \sqrt{\frac{\nz}{\nf}}\, \bar J^3_\su
\ee
have the property that there are no solutions to the null gauge constraints, because the charges under $J^3\tight-\bar J^3$ are integral, while the constraint generically forces some irrational relation between them.
However, there is a simple fix, which is to squash the $\bS^3$ and/or $AdS_3$ group manifold by adding (marginal) $J^3_\su\bar J^3_\su$, $J^3_\sl\bar J^3_\sl$ deformations of the WZW theory~\cite{Yang:1988bi,Hassan:1992gi,Giveon:1993ph}.  This deforms the spectrum of the current being gauged, and for appropriate choices of the squashing parameters the null constraints can be satisfied. 

The squashing deformation is an exactly marginal deformation of the sigma model.  Writing the $\sutwo$ sigma model Lagrangian as 
\be
\label{Lsu2 squashed}
\cL_\su = k_\su \bigg[ \partial\theta\bar\partial\theta + 
\begin{matrix} \Big(\partial\psi & \partial\phi \Big)  \end{matrix} \cdot \cE
\cdot\bigg(\begin{matrix}\bar\partial\psi \\ \bar\partial\phi \end{matrix}\bigg)  \bigg]
~~,~~~~
\cE = \bigg(\begin{matrix} \cos^2\theta & \; -\cos^2\theta \\ \cos^2\theta & ~~\sin^2\theta \end{matrix}\bigg)  ~~,
\ee
the squashing deforms the matrix $\cE=G+B$ to~\cite{Yang:1988bi,Hassan:1992gi,Giveon:1993ph}
\be
\label{Esquashed}
\cE(R_\su) = \frac{1}{\cos^2\theta+R_\su^2\sin^2\theta} \,
\bigg(\begin{matrix} R_\su^2\cos^2\theta & ~ -\cos^2\theta \\ \cos^2\theta & ~~\sin^2\theta \end{matrix}\bigg)
\ee
Reading off the metric and B-field from the parity even/odd terms in this Lagrangian, the deformed geometry admits ``left'' and ``right'' Killing vectors
\be
\label{SU squashed KVs}
\xi_{L,R}^\flat = \frac 1{R_\su}\partial_\psi \mp R_\su\,\partial_\phi  ~,
\ee
associated to the deformed $\uone$ currents
\begin{align}
\label{SU squashed currents}
\Jsq_\su &= k_\su R_\su\,\frac{\cos^2\theta\,\partial\psi - \sin^2\theta\,\partial\phi}{\cos^2\theta+R_\su^2\sin^2\theta}
\nn\\[.2cm]
\Jbsq_\su &= k_\su R_\su\,\frac{\cos^2\theta\,\bar\partial\psi + \sin^2\theta\,\bar\partial\phi}{\cos^2\theta+R_\su^2\sin^2\theta} 
\end{align}
which are the pull-backs to the worldsheet of the associated one-forms.
Note that the normalization of the two-point function of these currents is unaffected by the squashing deformation.
There is also a non-trivial dilaton in the squashed geometry
\be
e^{2\Phi} = \frac1R\cos^2\theta + R\sin^2\theta ~.
\ee

We can also squash the $AdS_3$.
The geometry is obtained by the substitutions $\theta\to i\rho$, $\psi\to\tau$, $\phi\to\sigma$ in the Lagrangian and currents~\eqref{Lsu2 squashed}-\eqref{SU squashed currents}, resulting in
%\emil{sign of B-field seems reversed relative to the above via analytic continuation, but agrees with the mathematica file}
\begin{align}
\label{Lsl2 squashed}
\cL_\sl &= k_\sl \bigg[ \partial\rho\bar\partial\rho + 
\begin{matrix} \Big(\partial\tau & \partial\sigma \Big)  \end{matrix} \cdot \cE(R_\sl)
\cdot\bigg(\begin{matrix}\bar\partial\tau \\ \bar\partial\sigma \end{matrix}\bigg)  \bigg]
\\[.2cm]
\cE(R_\sl) &= \frac{1}{\cosh^2\!\rho-R_\sl^2\sinh^2\!\rho} \,
\bigg(\begin{matrix} -R_\sl^2\cosh^2\!\rho & ~\cosh^2\!\rho \\ -\cosh^2\!\rho & ~\sinh^2\!\rho \end{matrix}\bigg)
\nn
\end{align}
and currents
\begin{align}
\label{SL squashed currents}
\Jsq_\sl &= k_\sl R_\sl\,\frac{\cosh^2\!\rho\,\partial\tau - \sinh^2\!\rho\,\partial\sigma}{\cosh^2\!\rho+R_\sl^2\sinh^2\!\rho}
\nn\\[.2cm]
\Jbsq_\sl &= k_\sl R_\sl\,\frac{\cosh^2\!\rho\,\bar\partial\tau + \sinh^2\!\rho\,\bar\partial\sigma}{\cosh^2\!\rho+R_\sl^2\sinh^2\!\rho} ~.
\end{align}
Note that the $\sltwo$ squashing parameter is restricted to $R_\sl<1$ in order to avoid a singularity in the geometry at some finite $\rho$; the $\sutwo$ squashing is unrestricted, though $R_\su \to 1/R_\su$ amounts to an interchange of the roles of $\phi$ and $\psi$ (the axial/vector duality of the $\sutwo$ WZW model).
The null gauge currents now take the form
\be
\label{AD null current}
\cJ = \Jsq_\sl - \sqrt{\frac{k_\sl}{k_\su}}\, \Jsq_\su
~~,~~~~
\bar\cJ = \Jbsq_\sl - \sqrt{\frac{k_\sl}{k_\su}}\, \Jbsq_\su 
\ee
up to an overall normalization.%
\footnote{In previous works~\cite{Martinec:2017ztd,Martinec:2018nco,Martinec:2019wzw,Martinec:2020gkv}, the convention was chosen to take opposite signs for the $\sutwo$ contributions to the two null currents, in order to match the literature for the effective geometry of supertubes.  Here we choose vector gauging instead, so as to have a symmetrical treatment of left- and right-movers on the worldsheet.}

%%%%%%%%%%%%%%%%%%%%%%%%%%%%%%%%%%%%%%
\subsection{Review of null gauging}
\label{sec:nullreview}

We now gauge these currents.
The kinetic terms in the sigma model action involve the covariant derivative on the target space coordinate fields $\varphi^i$
\be
\cD\varphi^i = \partial\varphi^i - \cA^a \xi_a^i(\varphi)
\ee
with gauge potentials $\cA^a$ for gauging the two null Killing vectors
\eq{
\xi_1 = \xi^\flat_{L,\sl} -\sqrt{\frac{k_\sl}{k_\su}}\,\xi^\flat_{L,\su}
~~,~~~~
\xi_2 = \xi^\flat_{R,\sl} -\sqrt{\frac{k_\sl}{k_\su}}\,\xi^\flat_{R,\su} ~~,
}
where $\xi^\flat_{L,R}$ are given \eg\ in~\eqref{SU squashed KVs} for $\sutwo$.
We have two independent gauge fields $(\cA^1,\cAb^1)$ and $(\cA^2,\cAb^2)$, one for each isometry being gauged; the sigma model kinetic term is 
\be
\label{gauged KE}
\cL_{\rm K} = \mathcal{D} \varphi^i \, G_{ij} \, \overline{\mathcal{D}} \varphi^j
\;=\; (\partial \varphi^i - \cA^a \:\! \xi_a^i ) \, G_{ij} \,  (\bar\partial \varphi^j - \cAb^a \:\! \xi_a^j) \;,
\ee
while the Wess-Zumino term is given in terms of target-space one-forms $\theta_a$, pulled back to the worldsheet:
\be
\label{eq:g-wz-term}
\cL_{\rm WZ} = B_{ij} \partial \varphi^i \bar\partial\varphi^j + \cA^a \theta_{a,i} \bar\partial\varphi^i + \cAb^a \theta_{a,i} \partial\varphi^i
+ \xi_{[a}^i \theta^{~}_{b],i} \cA^a \cAb^b \;.
\ee
For the pair of null Killing vectors $\xi_a$, the target-space one-forms $\theta_a$ are given by 
\be \label{eq:theta-xi}
\qquad\quad    \theta_{a} \;=\; (-1)^{a+1}  \xi_{a} \cdot d\varphi  \;\equiv\; (-1)^{a+1} \xi_a^i G_{ij} d\varphi^j \:\!      \qquad\quad (a=1,2)~;
\ee
this causes half the gauge field components to decouple. 
Due to these cancellations, the coefficient of the term quadratic in gauge fields ends up being proportional to the quantity
\be
\Sigma \;\equiv\; -\frac12\xi_1^i G_{ij} \xi_2^j   \,.
\ee 
For a consistent gauging, we have the conditions (see e.g.~\cite{Figueroa-OFarrill:2005vws}) 
\be
\imath_a H \;=\;  d\theta_a \,,\qquad \imath_a\theta_b \;=\; - \imath_b \theta_a   
\ee
(the latter condition amounts to the vanishing of the various anomalies among the currents).
Overall, the gauge field terms in the action reduce to
\be
-2\cA^2 \xi_2^i G_{ij} \bar\partial\varphi^j -2 \cAb^1 \xi_1^i G_{ij} \partial\varphi^j - 4 \cA^2 \cAb^1 \Sigma  ~.
\ee
In what follows, we denote $\cA\equiv \cA^2$, $\cAb\equiv \cAb^1$.

The worldsheet currents $\cJ$, $\bar{\cJ}$ are pull-backs of target-space one-forms as follows, 
\be
\label{eq:J-def}
\cJ \,\equiv\, - \theta_{1} \cdot \partial\varphi \,\equiv\, - \theta_{1,i} \:\! \partial\varphi^i \;, \qquad \bar\cJ \,\equiv\, \theta_{2} \cdot \bar\partial\varphi \,\equiv\, \theta_{2,i} \:\!\bar\partial\varphi^i \,.
\ee
Using \eqref{eq:theta-xi}, we can then rewrite the gauge terms as
\begin{align}
\label{eq:g-terms-gen}
2\cA \;\! \theta_{2,i} \:\!\bar\partial\varphi^i - 2\cAb \;\! \theta_{1,i} \:\! \partial\varphi^i - 4\cA \cAb \:\! \Sigma 
~\,\equiv\,~  2\cA \bar \cJ +  2\bar\cA \cJ -  4\cA \bar\cA \Sigma\,.
\end{align}
Integrating out the gauge fields leads to a term in the sigma model effective action
\be \label{eq:g-terms-int-out}
\frac{\cJ \bar\cJ}{\Sigma} ~=~ - \frac{1}{\Sigma} \big( \theta_{1} \cdot \partial\varphi  \big) \big( \theta_{2} \cdot \bar\partial\varphi \big) 
~=~ \frac{1}{\Sigma} \big( \xi_{1} \cdot \partial\varphi  \big) \big( \xi_{2} \cdot \bar\partial\varphi \big) \,,
\ee
where similarly $\xi_{1} \cdot \partial\varphi \equiv\xi_1^i G_{ij} \partial\varphi^j$.
Thus we see that the null gauging procedure in effect adds the terms~\eqref{eq:g-terms-int-out} to the sigma model Lagrangian.

%%%%%%%%%%%%%%%%%%%%%%%%%%%%%%%%%%%%%%
\subsection{Effective geometry, operator spectrum, etc.}
\label{sec:ADgeom}

Setting the gauge $\tau=\sigma=0$ and integrating out the gauge fields results in an effective transverse geometry
\begin{align}
\label{AD effective geom}
ds^2_\perp &= k_\sl \,d\rho^2 + k_\su\,d\theta^2
+\frac{R_\su^2 \cos^2\theta\,\cosh^2\rho}{R_\sl^2 \cos^2\theta\sinh^2\rho + R_\su^2 \sin^2\theta\,\cosh^2\rho}\, d\psi^2
\nn\\[.1cm]
&\hskip 2cm
+\frac{R_\sl^2\sin^2\theta\,\sinh^2\rho}{R_\sl^2 \cos^2\theta\sinh^2\rho + R_\su^2 \sin^2\theta\,\cosh^2\rho}\, d\phi^2
\\[.4cm]
\label{AD Beff}
B &= \Big[ \frac{R_\sl^2\, \cos^2\theta\sinh^2\rho}{R_\sl^2 \cos^2\theta\sinh^2\rho + R_\su^2 \sin^2\theta\,\cosh^2\rho} 
\Big] \,d\phi\wedge d\psi
\\[.4cm]
\label{AD Phieff}
e^{-2\Phi} &=  \frac{1}{g_s^2k_\su}\Big(
{\frac{R_\sl}{R_\su} \cos^2\theta\sinh^2\rho + \frac{R_\su}{R_\sl} \sin^2\theta\,\cosh^2\rho}
\Big) ~~.
\end{align}
The geometry is essentially a squashed version of the transverse space of the solution in the critical dimension, to which it reduces if we set $k_\sl=k_\su$ and $R_\sl=R_\su=1$.  Passing into the core of the geometry, one finds that the metric blows up at $\rho=0,\theta=0$ which is a circle parametrized by $\psi$, while the angular coordinate $\phi$ degenerates at this locus.  The interpretation given to this geometry in~\cite{Giveon:1999zm,Giveon:1999px,Aharony:2004xn} is that we have a discrete array of fivebranes on their Coulomb branch, uniformly distributed around the $\psi$ circle and wrapping the homological $\bS^2$ of the resolved conifold.  The homology cycle is so sub-stringy in size that it is invisible to perturbative strings and D-branes.  As a result, the dimensionality of target space of the worldsheet sigma model lies below the critical dimension, with the gradient of the asymptotically linear dilaton making up the difference in solving the beta function equations for the target space geometry. 

The spectrum of operators on the squashed $\bS^3$ has the form
\be
\label{Psihat squashed}
\Psihat_{j,m,\bar m}^{\eta,\bar\eta,w,\bar w} = \medhat \Lambda_{j,m,\bar m}^{\eta-w,\bar\eta-\bar w}
\exp\big[i\big( p^{~}_L \cY_\su + p^{~}_R\bar\cY_\su\big)\big] ~,
\ee
where $\medhat \Lambda$ is a ``super-parafermion" operator of the supersymmetric coset model $\frac\sutwo\uone$ and $\cY$ bosonizes the gauged $\uone$ current (for details and conventions, see Appendix~\ref{app:conventions}),
but now due to the deformation the $\cY$ exponentials have the momenta $p_{L,R}^{~}$ with $J^3_\su=\hf\sqrt {k_\su}\, p_L$ and conformal dimensions
\eq{
\label{pLR su squashed}
h[p^{~}_L]  &= \frac{p^{2}_L}{4} = \frac1{4k_\su}\Big( \frac{k_\su N+n}{R_\su} + R_\su\big( k_\su L+\ell \big) \Big)^2
\\[.2cm]
h[p^{~}_R]  &= \frac{p^{2}_R}{4} = \frac1{4k_\su}\Big( \frac{k_\su N+n}{R_\su} - R_\su\big( k_\su L+\ell \big) \Big)^2
}
with 
\begin{align}
\label{LR vs AV}
2(m+\eta-w) = n+\ell ~~&,~~~~  w = N+L
\\
2(\bar m+\bar\eta-\bar w) = n-\ell ~~&,~~~~\bar w=N-L 
\quad,\qquad n,\ell,N,L\in\bZ
\nn
\end{align}
(note that for $R_\su=1$ these expressions reduce to the usual parafermion decomposition of current algebra representations, see Appendix~\ref{app:conventions}).  
The operator spectrum for the $\sltwo$ factor similarly takes the form
\be
\Phihat_{j,m,\bar m}^{\eta,\bar\eta,w,\bar w} = \medhat V_{j,m,\bar m}^{\eta+w,\bar\eta+\bar w}
\exp\big[i\big( p^{~}_L \cY_\sl + p^{~}_R\bar\cY_\sl\big)\big] ~,
\ee
with the analogous expressions for the left/right momenta $p_{L,R}$, except that the axial winding $L_\sl=0$ since we are working on the universal cover $AdS_3$ of $\sltwo$.

Physical states must satisfy the null gauge constraints $\cJ=0=\bar\cJ$, or
\eq{
\label{AD null constraints}
p_L^\sl -  p_L^\su = 0
~~,~~~~
p_R^\sl -  p_R^\su = 0 ~.
}
For generic $R_\su,R_\sl$ there are no solutions to these constraints.  We choose 
\be
\frac{R_\sl}{\sqrt{k_\sl}} = 
\frac{R_\su}{\sqrt{k_\su}} \equiv R 
\ee
which makes the orbits of the spatial Killing vector closed (\ie\ the isometry generated by $\cJ-\bar\cJ$), and thus allows the constraints to be solved.
The null constraints~\eqref{AD null constraints} become
\be
\label{nullcon2}
\frac{n_\sl+k_\sl N_\sl}{{k_\sl}\,R} \pm \ell_\sl \,R = 
\frac{n_\su+k_\su N_\su}{{k_\su}\,R} \pm \big(\ell_\su+k_\su L_\su\big)\,R
\ee
Note that the overall scale $R$ drops out of the effective geometry; its arbitrariness forces the various powers of $R$ in the above constraints to be satisfied separately.  Thus we conclude
\be
\frac{n_\sl+k_\sl N_\sl}{k_\sl} = \frac{n_\su+k_\su N_\su}{k_\su}
~~,~~~~
\ell_\sl = \ell_\su+k_\su L_\su
\ee

In general, vertex operators are composed of center-of-mass operators schematically of the form
\be
\Phihat_\sl \, \Psihat_\su \, \exp\big[ip_\mu X^\mu\big]
\ee
from each of the factors $\sltwo\times\sutwo\times\bR^{3,1}$, times a polynomial in the various currents and fermions, and derivatives thereof.
Physical state vertex operators must solve the null gauge constraints~\eqref{AD null constraints}, and the Virasoro constraints  which impose (among other things) the mass shell condition and level-matching constraints
\eq{
\label{AD Vir}
0 &= -\frac{\jsl(\jsl-1)-\msl^2}{\nz}-\frac14 \big(p_L^\sl\big)^2
+\frac{\jsu(\jsu+1)-\msu^2}{\nf}+\frac14 \big(p_L^\su\big)^2 + \frac14 p_\mu p^\mu + N_L-\half
\\[.2cm]
0 &= -\frac{\jsl(\jsl-1)-\bmsl^2}{\nz}-\frac14 \big(p_R^\sl\big)^2
+\frac{\jsu(\jsu+1)-\bmsu^2}{\nf}+\frac14 \big(p_R^\su\big)^2 + \frac14 p_\mu p^\mu + N_R-\half  ~.
}
The allowed $\jsu$ representations for normalizable excitations lie in the range
\be
\label{jsu range}
0 \le \jsu \le \frac{\nf}{2} -1 ~.
\ee
The spectrum of $\jsl$ consists of discrete series representations $\cD^\pm_{\jsl}$ with
\be
\label{jsl range}
\half < \jsl < \frac{\nz+1}{2}
\ee
describing modes bound to the bottom of the throat, and continuous series representations $\cC_{\jsl}$ with $\jsl\in\hf+i\lambda$, $\lambda\in \bR$, describing a continuum of scattering states propagating up and down the linear dilaton throat.  

Of particular interest are the BPS operators, which come from highest/lowest weight operators of the current algebra and have $p^\mu=0$.  These are schematically of the same form as for the critical dimension~\cite{Martinec:2020gkv}
\begin{align}
\label{BPS vertexops}
\cV_j^{-} &= e^{-\varphi} \Phi^{\sl}_{\jsl} \big( \psi_{\su} \Psi_{\jsu}^{\su}\big)_{\jsu+1}  \nn\\[.1cm]
\cV_j^{+} &=e^{-\varphi} \big(\psi_{\sl} \Phi_{\jsl}^{\sl}\big)_{\jsl-1} \Psi^{\su}_{\jsu}  \\[.1cm]
\cS_j^{\alpha} & = e^{-(\varphi+\tilde\varphi)/2} \bigl(S^{\alpha} \Phi_{\jsl}^{\sl} \Psi_{\jsu}^{\su}\bigr)_{\jsl-1/2,\jsu+1/2}  ~~, \nonumber
\end{align}
where the $\Phi$, $\Psi$ are in highest weight states of their respective bosonic current algebras, and the subscripts on parentheses indicate the projection of the tensor product of the spin of the center-of-mass vertex operator and the spin one fermion $\psi$ (or spin 1/2 spin field $S^\alpha$) onto the indicated total spin.  In the critical theory, the constraints impose $\jsl=\jsu+1$, but here the Virasoro and null constraints impose a different choice.

More precisely, in the models we consider here, the $\sutwo$ and $\sltwo$ are squashed by a $J^3\bar J^3$ deformation, but this destroys the factorization of the operators into a fermion contribution and a bosonic current algebra representation.  Nevertheless, we can employ the parafermion representation sketched in Appendix~\ref{app:conventions} and follow the squashing deformation of the above vertex operators away from the $\sutwo$ point specified in~\eqref{Psihat squashed}-\eqref{LR vs AV}.  The particular fermion polarizations involved in BPS states correspond to turning on the parameter $\eta$ of~\eqref{LR vs AV}.

The parafermion decomposition of the operators~\eqref{BPS vertexops} takes the form (suppressing the ghost factors to reduce clutter)
\begin{align}
\label{AD chiralops}
&\medhat V_{\jsl,\jsl,\jsl}^{\hat\eta_\sl,\hat\eta_\sl} \, \medhat\Lambda_{\jsu,\jsu,\jsu}^{\tilde\eta_\su,\tilde\eta_\su} 
\exp\Big[ i p_\sl(\cY_\sl+\bar\cY_\sl) + i p_\su(\cY_\su+\bar\cY_\su) \Big]
\nn\\[.2cm]
&\medhat V_{\jsl,\jsl,\jsl}^{\hat\eta_\sl,\hat\eta_\sl} = V_{\jsl,\jsl,\jsl}^{~} \exp\Bigl[ i\sqrt{\coeff{8}{k_\sl(k_\sl+2)}} \,\Bigr(\jsl+\coeff{k_\sl+2}{2}\hat\eta_\sl\Bigr)\big(\cH_\sl+\bar\cH_\sl\big)\Bigr]
\\[.2cm]
&\medhat \Lambda_{\jsu,\jsu,\jsu}^{\tilde\eta_\su,\tilde\eta_\su} = \Lambda_{\jsu,\jsu,\jsu}^{~} \exp\Bigl[ i\sqrt{\coeff{8}{k_\su(k_\su-2)}} \,\Bigr(-\jsu+\coeff{k_\su-2}{2}\tilde\eta_\su\Bigr)\big(\cH_\su+\bar\cH_\su\big)\Bigr]
\nn
\end{align}
where $\hat\eta\equiv\eta\tight+w$, $\tilde\eta\equiv\eta\tight-w$; $\cH_\sl,\cH_\su$ bosonize the $\cR$-symmetry currents of the superparafermions; $V$ and $\Lambda$ are parafermions of the underlying bosonic coset models;
and the quantum numbers that satisfy the Virasoro and null constraints for \eg\ $\cV^+$ are given by
\begin{align}
\label{AD V+ quantum nums}
\jsl = 1 + \frac{2\jsu}{k_\su+2}
~~,~~~~
\eta_\sl \tight= \bar\eta_\sl \tight= -1
~~,~~~~
\eta_\su \tight= \bar\eta_\su \tight= w_\su \tight= \bar w_\su \tight= w_\sl \tight= 0
\end{align}
(from which one infers $p_{L,R}$ using~\eqref{LR vs AV}).  
The $\sutwo$ spin takes the allowed values $\jsu\in 0,\hf,\ldots,\hf k_\su\tight-1$.
Similarly, for $\cV^-$ one has
\begin{align}
\label{AD V- quantum nums}
\jsl = \frac{2\jsu+2}{k_\su+2}
~~,~~~~
\eta_\su \tight= \bar\eta_\su \tight =1
~~,~~~~
\eta_\sl \tight= \bar\eta_\sl \tight= w_\su \tight= \bar w_\su \tight= w_\sl \tight= 0
\end{align}
and for $\cS^\pm$ one has either of these values of $j_\sl$, as well as
\begin{align}
\label{AD S quantum nums}
\eta_\su \tight= \bar\eta_\su \tight =\half
~~,~~~~
\eta_\sl \tight= \bar\eta_\sl \tight= -\half
~~,~~~~
w_\su \tight= \bar w_\su \tight= w_\sl \tight= 0  ~~.
\end{align}
These operators generate marginal deformations which move the fivebranes along their Coulomb branch, away from the circular array.  This spectrum agrees with what one obtains from the coset orbifold construction of%
~\cite{Giveon:1999zm,Giveon:1999px,Aharony:2004xn}, as we review below.  
Note also that these supergraviton states are equivalent to ``winding tachyon'' states in the parent WZW model that survive the gauge projection via the generalized FZZ duality described in~\cite{Giveon:2016dxe,Martinec:2020gkv}.

Note that the unitarity ranges~\eqref{jsu range}, \eqref{jsl range} restrict the solutions of~\eqref{AD V+ quantum nums}.
For $\cV^\pm$ the allowed ranges for generic $k_\su$ are
\be
\label{restricted ranges}
\cV^+ \,:\quad 0\le j_\su < \frac{k_\su-2}4
\quad,\qquad
\cV^- \,:\quad \frac{k_\su-2}4 \le j_\su < \frac{k_\su-2}2 ~.
\ee
Putting everything together, one has the following 1/2-BPS vertex operators and their equivalences, restricting attention to the right-moving component for simplicity, and denoting by $j',j$ the values of $\jsl,\jsu$ given in~\eqref{AD V+ quantum nums}, \eqref{AD V- quantum nums}:
%
%%%%%%%%%%%%%%%%%%%%%%%%%%%%%%%
%
\begin{centermath}
\arraycolsep=1.4pt\def\arraystretch{1.2}
\begin{array}{|c|c|c|c||c|c|c|c||c||c|c|c|c|}
\hline
j_\sl &  m_\sl & \eta_\sl & w_\sl & j_\su &  m_\su &  \eta_\su &  w_\su & w_y & \cH_\sl & \cH_\su & \cY_\sl & \cY_\su \\[2pt]
\hline
\hline
j' & j' & -1 & 0 & j & j & 0 & 0 & w & \frac{2j}{k_\sl} \tight-1 & -\frac{2j}{\ksl} & \frac{2j}{\ksu} & \frac{2j}{\ksu} \\[2pt]
\hline
\frac{\hatksl}2\tight-j' & j'\tight-\frac{\hatksl}2& -1 & 1 & j & j & 0 & 0 & w & \frac{2j}{\ksl} \tight-1 & -\frac{2j}{\ksl} & \frac{2j}{\ksu} & \frac{2j}{\ksu} \\[2pt]
\hline
\frac{\hatksl}2\tight-j' & j'\tight-\frac{\hatksl}2 & 0 & 0 & j & j & -1 & -1 & w\tight-k & \frac{2j}{\ksl} \tight-1& -\frac{2j}{\ksl} & \frac{2j}{\ksu} \tight-1 & \frac{2j}{\ksu} \tight-1 \\[2pt]
\hline
\frac{\hatksl}2\tight-j' & j'\tight-\frac{\hatksl}2 & 0 & 0 & \frac{\tilksu}2\tight-j & j\tight-\frac{\tilksu}2 & -1 & 0 & w\tight-k & \frac{2j}{\ksl} \tight-1 & -\frac{2j}{\ksl} & \frac{2j}{\ksu} \tight-1 & \frac{2j}{\ksu} \tight-1 \\[2pt]
\hline
\hline
j' & j' & 0 & 0 & j & j & 1 & 0 & w & \frac{2j}{\ksl} & 1\tight-\frac{2j}{\ksl} & \frac{2(j\tight+1)}{\ksu} & \frac{2(j\tight+1)}{\ksu} \\[2pt]
\hline
j' & j' & 0 & 0 & \frac\tilksu 2\tight-j & j\tight-\frac\tilksu 2 & 1 & 1 & w & \frac{2j}{\ksl} & 1\tight-\frac{2j}{\ksl} & \frac{2(j\tight+1)}\ksu & \frac{2(j\tight+1)}\ksu \\[2pt]
\hline
j' & j' & 1 & -1 & \frac{\tilksu}2\tight-j & j\tight-\frac{\tilksu}2 & 0 & 0 & w\tight-k & \frac{2j}{\ksl} & 1\tight-\frac{2j}{\ksl} & \frac{2(j\tight+1)}{\ksu}\tight-1 & \frac{2(j\tight+1)}{\ksu}\tight-1 \\[2pt]
\hline
\frac{\hatksl}2\tight-j' & j'\tight-\frac{\hatksl}2 & 1 & 0 & \frac{\tilksu}2\tight-j & j\tight-\frac{\tilksu}2 & 0 & 0 & w\tight-k & \frac{2j}{\ksl} & 1\tight-\frac{2j}{\ksl} & \frac{2(j\tight+1)}{\ksu}\tight-1 & \frac{2(j\tight+1)}{\ksu}\tight-1 \\[2pt]
\hline
\end{array}
\end{centermath}
{{Table 1. }\textit{\small Equivalences among vertex operators.  The first block of four are all equivalent representatives of the same state; similarly, the entries in the second block of four are all equivalent to one another.  The $w_y$ column refers to the supertube construction of section~\ref{sec:supertubes}; for the purposes of this section one can set $w_y=k=0$.  We also define $\hatksl=k_\sl+2,\tilksu=k_\su-2$.}}
%
%%%%%%%%%%%%%%%%%%%%%%%%%%%%%%%

\vskip .5cm

\noindent
The passage from the first to the second line of the table involves the use of FZZ duality~\cite{FZZref,Giveon:1999px,Giveon:1999tq} on $\sltwo$, an operator identification in the loop group (see~\cite{Martinec:2020gkv} for a discussion in the present context).  Gauge spectral flow takes one from the second line to the third, and the third line is related to the fourth by FZZ duality in $\sutwo$.  The second block of four rows are similarly related (in reverse order).

%%%%%%%%%%%%%%%%%%%%%%%%%%%%%%%%%%%%%%
%%%%%%%%%%%%%%%%%%%%%%%%%%%%%%%%%%%%%%

\subsection{Comparison to coset orbifold description}
\label{sec:AD coset orbifold}

The original construction of~\cite{Sfetsos:1998xd,Giveon:1999px} presents the worldsheet theory of fivebranes on the Coulomb branch as an orbifold of a tensor product of coset models
\be
\bigg(\frac{\sltwo}{\uone}\times \frac{\sutwo}{\uone}\bigg)\Big/\bZ_{k_\su} ~.
\ee
This model has the same spectrum as the null gauged model above.  For instance, consider the BPS spectrum.  The chiral ring of the coset tensor product consists of the fields 
\be
\label{AD chiralops2}
\medhat V_{\jsl,\jsl,\jsl}^{0,0} \, \medhat\Lambda_{\jsu,\jsu,\jsu}^{0,0} ~;
\ee
the Virasoro constraints impose
\be
\frac{j_\sl}{k_\sl}  + \frac{j_\su}{k_\su} = \half  ~,
\ee
which determines $j_\sl$ in terms of $j_\su$.  The operator one arrives at is related to~\eqref{AD chiralops} by FZZ duality (see~\cite{Martinec:2020gkv} for a discussion).
The $\Neqtwo$ $\cR$-charges of these operators are $\frac{2j_\su}{k_\su}$ and $\frac{2j_\sl}{k_\sl} = 1-\frac{2j_\su}{k_\su}$, and so survive the $\bZ_{k_\su}$ orbifold projection, which requires that the sum of the $\cR$-charges is an integer.  This integer charge property allows the possibility of a chiral GSO projection by ensuring an absence of branch cuts in correlation functions.

The null gauging construction arrives at this spectrum in a slightly different way.  Instead of the coset, one has its parent (squashed) WZW model; and instead of the orbifold, one has the null gauge constraints.  The quantum numbers of the $\cY$ exponential are correlated to those of the $\cH$ exponential in~\eqref{AD chiralops}; the gauge constraint~\eqref{AD null constraints} relates the $\cY$ charges of $\sltwo$ and $\sutwo$, and thus the $\cH$ charges.  In the end, one arrives at the same physical spectrum.

Worldsheet dualities provide an alternative proposed interpretation for this background~\cite{Giveon:1999zm,Giveon:1999px}.  The $\frac{\sutwo}{\uone}$ coset model at level $n$ is quantum equivalent to the $\Neqtwo$ Landau-Ginsburg theory~\cite{Kastor:1988ef,Martinec:1988zu,Vafa:1988uu}, and the $\frac{\sltwo}{\uone}$ coset is quantum equivalent to $\Neqtwo$ Liouville theory~\cite{FZZref,Giveon:1999px}.  The Landau-Ginsburg interpretation of the coset orbifold~\cite{Ooguri:1995wj} is that the $\Neqtwo$ superpotential $\cW(\sfZ)=\sfZ^{k_\su}$ describes a slightly resolved Calabi-Yau hypersurface singularity
\be
\sfZ^{k_\su}+\sfZ_{\!1}^{\,2}+\sfZ_{\!2}^{\,2} +\sfZ_{\!3}^{\,2}  = \lambda_0
\ee
in $\bC^4$.  The worldsheet duality that relates the hypersurface singularity description and the fivebrane description of the background amounts to mirror symmetry~\cite{Hori:2001ax}.

The role of the Liouville field is to dress the resolution parameter $\lambda_0$ so that it is the coupling of a marginal operator on the worldsheet.  In fact, all the chiral operators~\eqref{AD chiralops2} generate deformations of the singularity~-- the operators $\medhat\Lambda_{\jsu,\jsu,\jsu}^{0,0}$, with $j_\su\tight\le \frac{k_\su-2}{2}$, are the powers of the generator $\sfZ=\medhat\Lambda_{\half\half\half}^{0,0} $ of the Landau-Ginsburg chiral ring, and again the role of the attached $\Neqtwo$ Liouville field is to dress this deformation to make a marginal worldsheet operator.  
Note that due to the restricted ranges~\eqref{restricted ranges} of the allowed spins, the array of deformations of the fivebrane array is more limited in the Argyres-Douglas background than in the critical dimension.

In the null gauging approach, this whole construction lifts up to the parent tensor product of (squashed) WZW models~\cite{Martinec:2020gkv}.  The duality between the coset models and their Landau-Ginsburg duals is embodied in the operator equivalences of Table~1.  Here one can understand the origin of FZZ duality as being a consequence of operator identifications related to spectral flow; for details, see~\cite{Martinec:2020gkv}.

An alternative interpretation~\cite{Martinec:2020gkv} of the FZZ dual picture posits that the superpotential codes the near-source structure of the fivebrane configuration, in that the zeroes of the superpotential code the fivebrane locations in the $x^1\!\tight+ix^2$ plane.  The Liouville plus Landau-Ginsburg superpotential factorizes as
\eq{
\cW = \sfZ^{k_\su} - \lambda_0\, e^{k_\su  \sfX} 
= \prod_{\ell=1}^{k_\su} \Big( Z - \mu_\ell \,e^{ \sfX} \Big)
}
where $\mu_\ell = (-\lambda_0)^{1/k_\su} e^{2\pi i \ell/k_\su}$.
The FZZ duals of the $\cV^\pm$ operators~\eqref{AD V+ quantum nums}, \eqref{AD V- quantum nums} are the operators 
\eq{
\lambda_n\, \sfZ^n  e^{(k_\su-n)\sfX}
}
that deform the superpotential and hence the locations of the fivebranes.  This alternative interpretation provides a consistent picture of the near-source structure of supertubes~\cite{Martinec:2020gkv}, as we will see in section~\ref{sec:supertubes}.

%%%%%%%%%%%%%%%%%%%%%%%%%%%%%%%%%%%%%%
%%%%%%%%%%%%%%%%%%%%%%%%%%%%%%%%%%%%%%

\subsection{2d version}
\label{sec:2d version}

A similar story holds for fivebranes wrapping the vanishing cycles of a slightly resolved Calabi-Yau fourfold singularity.  One simply replaces the longitudinal $\IR^{3,1}$ with $\IR^{1,1}$ in the above.  The effective geometry (to the extent that it is reliable) is again given by~\eqref{AD effective geom}-\eqref{AD Phieff}, but now the current algebra levels are related by
\be
k_\sl = \frac{k_\su}{k_\su+1} ~~.
\ee
In particular, the $\sltwo$ level is always smaller than one~-- below the correspondence point for linear dilaton backgrounds~\cite{Giveon:2005mi}.  The variation of the background is large relative to the string scale $\alpha'$, and the effective geometry can be a poor description of the effective dynamics.  The $\sltwo/\uone$ coset exhibits a strong-weak coupling duality to $\Neqtwo$ Liouville theory, and for $k_\sl<1$ it is the Liouville description that is more weakly coupled.  For instance, it is argued in~\cite{Giveon:2005mi} that the black hole solution in linear dilaton and $AdS_3$ backgrounds is non-normalizable for $k_\sl<1$, so that instead of the high-energy spectrum being dominated by black hole microstates, it  instead consists of the Hagedorn spectrum of fundamental strings.  We will return to this issue in the concluding section.

The BPS spectrum for these models is quite limited~\cite{Giveon:1999zm}.  The range of bosonic discrete series representations~\eqref{jsl range} has an upper bound less than one, but the chiral operators of the form $\cV^+$ have $\sltwo$ spins
\be
j_\sl = 1+\frac{2j_\su}{2(k_\su+1)}
\ee
which all lie outside the unitary range for $j_\su$ in the allowed range~\eqref{jsu range}.  Similarly, the operators of the form $\cV^-$ have spins 
\be
j_\sl = \frac{2j_\su+2}{2(k_\su+1)}
\ee
that all lie in the range $j_\sl<\half$ and so are also outside the allowed range.  One can construct a set of RR vertex operators $\cS_j$ with
\begin{align}
\label{2d S quantum nums}
\jsl = \half+ \frac{2\jsu+1}{2(k_\su+1)}
~~,~~~~
\eta_\su \tight= \bar\eta_\su \tight =\half
~~,~~~~
\eta_\sl \tight= \bar\eta_\sl \tight= -\half
~~,~~~~
w_\su \tight= \bar w_\su \tight= w_\sl \tight= 0  
\end{align}
whose $\sltwo$ spins lie in the allowed range~\eqref{jsl range}.

%%%%%%%%%%%%%%%%%%%%%%%%%%%%%%%%%%%%%%
%%%%%%%%%%%%%%%%%%%%%%%%%%%%%%%%%%%%%%

\section{Fivebrane intersections over $\bR^{1,1}$ and $\bR^{2,1}$}
\label{sec:1d-2d intersection}

The above technology can also be used to describe throats of stacks of intersecting fivebranes, by including a separate $\sutwo$ factor to account for the $H$ flux through the angular 3-cycle transverse to each stack.  In this section we describe some elementary examples of this sort, where the fivebranes intersect in codimension eight; and then move on to more complicated constructions for codimension four intersections in the next section.

%%%%%%%%%%%%%%%%%%%%%%%%%%%%%%%%%%%%%%
%%%%%%%%%%%%%%%%%%%%%%%%%%%%%%%%%%%%%%

\subsection{Intersections over $\bR^{1,1}$}
\label{sec:1d intersection}

Two stacks of $\nfive,\nfivetil$ fivebranes intersecting over $\bR^{1,1}$ source a geometry that is simply two copies of that of a single stack.  Then we simply take two copies of the above null quotient, one for the space transverse to the first stack and one for the space transverse to the other
\be
\bigg(\frac{\sltwo \times \sutwo}{U(1)_\ell\times U(1)_r}\bigg) \times \bigg(\frac{\widetilde{SL}(2,\bR) \times \sutwotil}{\widetilde{U}(1)_\ell\times \widetilde{U}(1)_r} \bigg) \times \bR^{1,1}  
\ee
where we will use tilde's to denote the second group factor.  The geometry for this quotient is thus two copies of the Coulomb branch fivebrane geometry
\begin{align}
\label{R11 intersection geom}
ds^2 &=  -du\,dv  
+ \nfive\Bigl[d\rho^2 + d\theta^2 + \frac{1}{\Sigma_0}\Bigl(
{\cosh}^2\!\rho\cos^2\!\theta \,d\phi^2 + {\sinh}^2\!\rho\sin^2\!\theta \,d\psi^2\Bigr)\Bigr]
\nn\\[.2cm]
&\hskip 1.7cm
+ \nfivetil\Bigl[d\tilde\rho^2 + d\thetat^2 + \frac{1}{\widetilde\Sigma_0}\Bigl(
{\cosh}^2\!\tilde\rho\cos^2\!\thetat \,d\phit^2 + {\sinh}^2\!\tilde\rho\sin^2\!\thetat \,d\psit^2\Bigr)\Bigr]
\nn\\[.2cm]
B   &= \frac{\nfive \cos^2\theta \sinh^2\!\rho}{\Sigma_0} \, d\phi \wedge d\psi
+ \frac{\nfivetil \cos^2\thetat \sinh^2\!\tilde\rho}{\widetilde\Sigma_0} \, d\phit \wedge d\psit
\nn\\[.2cm]
e^{-2\Phi} &= \frac{\Sigma_0\widetilde\Sigma_0}{\gstrsq\nfive\nfivetil}  ~~, \qquad 
\Sigma_0 \;\equiv\; {\sinh^2\!\rho\,\cos^2\!\theta + \cosh^2\!\rho\,\sin^2\!\theta} ~.
\end{align}
The spectrum and other properties of this background are a straightforward application of the results of~\cite{Martinec:2017ztd,Martinec:2018nco,Martinec:2019wzw,Martinec:2020gkv}.

%%%%%%%%%%%%%%%%%%%%%%%%%%%%%%%%%%%%%%
%%%%%%%%%%%%%%%%%%%%%%%%%%%%%%%%%%%%%%

\subsection{Intersections over $\bR^{2,1}$}
\label{sec:2d intersection}

One can describe two stacks of fivebranes intersecting over $\bR^{2,1}$ using a null quotient
\be
\label{2+1 intersection}
\bigg(\frac{\sltwo\times\sutwo\times\widetilde{SU}(2)}{\uone_\ell\times\uone_r}\bigg)  \times \bR^{2,1}
\ee
where the $\sltwo$ and $\sutwo$ levels are related via
\be
\label{n0 value}
\nz = \frac{k_\su k_\sutil}{k_\su+k_\sutil} ~.
\ee
This background was considered in~\cite{Itzhaki:2005tu} as the very near-source limit of intersecting fivebranes in codimension eight.  When the fivebranes are {\it not} separated in their transverse space, the radial direction in each fivebrane throat is described on the worldsheet by a free field with linear dilaton~\cite{Callan:1991at}.  The two radial directions $\rho,\rhotil$ tranverse to each stack have linear dilatons with slopes
\be
Q = \sqrt{1/k_\su}
~~,~~~~
\tilde Q = \sqrt{1/k_\sutil} ~.
\ee
One then forms the linear combinations 
\be
\hat Q\vartheta = \tilde Q \rho - Q \rhotil
~~,~~~~
\hat Q\varrho = Q\rho + \tilde Q \rhotil ~;
\ee
note that $\vartheta$ codes the angular direction in this two-dimensional radial space, and has no linear dilaton.  It was conjectured by~\cite{Itzhaki:2005tu} that the near-horizon symmetry is enhanced from 1+1d to 2+1d (super)Poincar\'e invariance.  One then supposes that the desingularization of this throat along the Coulomb branch amounts to the embedding of the linear dilaton direction into $\sltwo$ and getting rid of the additional two dimensions via null gauging.

Separating each stack of fivebranes onto their Coulomb branch, we can gauge the null currents
\be
\label{R21 gauging}
\cJ = a \, J^3_\sl  - \Jsq_\su - \Jsq_\sutil
~~~,~~~~~
\bar \cJ = a \,J^3_\sl - \Jbsq_\su - \Jbsq_\sutil 
~~,~~~~
a = \frac{\nf+\nft}{\sqrt{\nf\nft}}
\ee
where we have squashed the $\sutwo$ group manifolds but do not squash $\sltwo$.  The squashed currents are given by~\eqref{SU squashed currents}.
Setting the gauge $\tau=\sigma=0$, the gauge action becomes
\begin{align}
\cL_{\rm gauge} &=  2\cA \bar\cJ + 2\bar\cA \cJ + 4\Sigma\, \cA\bar\cA 
\nn\\[.1cm]
2\Sigma &= (k_\su+k_\sutil)\cosh2\rho - k_\su\,f[\theta,R_\su] - k_\sutil\,f[\thetat,R_\sutil]
\end{align}
where we have defined
\be
\label{fdefs}
\ff(x,R) = \frac{f_-(x,R)}{f_+(x,R)}
~~,~~~~
f_\pm(x,R) = \frac1R \cos^2x  \pm R\, \sin^2x
~~.
\ee
Integrating out the gauge field results in contribution $\cJ\bar\cJ/\Sigma$ to the effiective action, and thus a singularity where $\Sigma=0$, which occurs at $\rho\tight=\theta\tight=\thetat\tight=0$.  This locus is the $\bT^2$ parametrized by $\psi,\psit$.  We expect $k_\su^{~} k_\sutil^{~}$ intersection points of the two stacks of fivebranes to be spread in a uniform array along this torus.  Presumably, this intersection structure should be visible to D-brane probes as in~\cite{Elitzur:2000pq,Israel:2005fn,Martinec:2019wzw}.
Asymptotically, the geometry approaches a linear dilaton throat with angular geometry squashed $\bS^3\times\widetilde\bS^3$, which one can read off from~\eqref{Lsu2 squashed}-\eqref{Esquashed}
\eq{\label{R12SU2 Squashed}
ds^2 \sim  k_\sl\, d\rho^2 &+
k_\su\left( d\theta^2+\frac{ \tan^2\theta}{1+R_\su^2\tan^2\theta}d\phi^2+
\frac{R_\su^2}{1+R_\su^2\tan^2\theta}d\psi^2\right)
\\[.2cm]
&+k_\sutil\left(d\thetat^2+\frac{\tan^2\thetat}{1+R_\sutil^2 \tan^2\thetat}d\phit^2+ \frac{R_\sutil^2}{1+R_\sutil^2 \tan^2\thetat}d\psit^2\right)~.
}

Once can again work out the spectrum of BPS states and the Coulomb branch moduli.  The squashed $\sutwo$ factor has a $J^\flat,\bar J^\flat$ spectrum~\eqref{pLR su squashed}
\be
\sqrt{k_\su}\, p_{L,R} = \frac{n_\su + k_\su N_\su}{2R_\su} \pm \frac{ R_\su}2 \big(\ell_\su + k_\su L_\su\big)
\ee
and similarly for $\sutwotil$;
and so with $R_\su\tight=R_\sutil\tight=a^{-1}$, highest weight states in all three factors, and a single fermion in one of the three factors carrying the supergraviton polarization analogous to $\cV^\pm$ of~\eqref{BPS vertexops} satisfy the null constraints $\cJ=\bar\cJ=0$ provided
\be
a\, \big(m_\sl+\eta_\sl \big) - \frac{n_\su}{2R_\su} - \frac{n_\sutil}{2R_\sutil} = a\, \big(\bar m_\sl+\bar\eta_\sl \big) - \frac{\bar n_\su}{2R_\su} - \frac{\bar n_\sutil}{2R_\sutil}  = 0
\ee
(also subject to the unitarity bounds~\eqref{jsu range}, \eqref{jsl range}; 
these are 1/2-BPS in spacetime.

%%%%%%%%%%%%%%%%%%%%%%%%%%%%%%%%%%%%%%
%%%%%%%%%%%%%%%%%%%%%%%%%%%%%%%%%%%%%%

\section{Intersections over $\bR^{3,1}$}
\label{sec:3d intersection}

The transverse space to the intersecting fivebranes can be reduced by one further dimension if one gauges an additional spacelike isometry; consequently, one can expand the longitudinal directions of the intersection to 3+1d.   We will consider two possibilities.  First, we can begin with the above supersymmetric coset~\eqref{2+1 intersection} and gauge a third, spatial isometry.  It turns out that the conditions of anomaly cancellation force the additional isometry to have fixed points; these fixed points will occur all up and down the throat (\ie\ for all $\rho$).  These fixed points naively lead to a diverging dilaton at a particular locus in the angular space of the fivebranes; however we will see that the naively diverging dilaton is an artifact of the duality frame, and a different choice of duality frame removes the singularity, while unfortunately introducing a similar singularity in another corner of the angular space.  In this sense, the angular space is not a conventional geometry but rather a {\it T-fold}~-- a space whose transition functions are not simply coordinate transformations, but may in addition involve T-duality transformations.  The simplest example of this phenomenon is in fact the coset model $\frac{\sutwo}{\uone}$~\cite{Kiritsis:1991zt,Giveon:1994fu,Maldacena:2001ky}, so it is not surprising to see the same phenomenon cropping up here.

The second approach tries to embed the additional $\uone$ isometry in $\sutwo\times\sutwotil$ so that the orbits don't have fixed points.  The resulting angular coset was first explored in~\cite{Guadagnini:1987ty} and considered in the present context in~\cite{PandoZayas:2000he}; this angular geometry resembles that of the Sasaki-Einstein manifolds $T^{p,q}$ but with some important differences.  

Here we will use the null gauging formalism from~\cite{Martinec:2019wzw} in order to study NS5-branes wrapping cycles in the spaces $T^{p,q}$. This is constructed by starting with a WZW model on the smooth 11+2d space 
\be
\big(\sltwo\times \sutwo \times \sutwotil \big) \times \IR^{3,1} ~,
\ee
and then gauging two null and one spacelike Killing vectors:
\be
\bigg(\frac{\sltwo\times \sutwo \times \sutwotil}{U(1)_\ell\times U(1)_r\times U(1)_s} \bigg) \times \IR^{3,1}  ~.
\ee
%Here we use the notation $\widetilde{SU}(2)$ simply to denote the second $SU(2)$ factor. 
%
It remains to choose the embedding of the left and right null isometries, as well as the spatial isometry $\uone_s$.  
%It is convenient to work in Euler angle coordinates~\eqref{eq:su2-param}, \eqref{eq:sl2-param} for $\sutwo$ and $\sltwo$, respectively.

%%%%%%%%%%%%%%%%%%%%%%%%%%%%%%%%%%%%%%%%%%%
%%%%%%%%%%%%%%%%%%%%%%%%%%%%%%%%%%%%%%%%%%%

\subsection{Option 1: Symmetric gauging and T-folds}

If the spatial isometry has fixed points in $\sutwo \times \widetilde{SU}(2)$, the angular space transverse to the fivebranes will not be a conventional geometry~-- in gauged WZW models, the projected metric gets a warp factor (and the dilaton varies) according to the size of the gauge orbit, resulting in an apparent singularity where the orbits degenerate.  

The simplest example of this phenomenon is the coset $\frac{\sutwo}{\uone}$, sometimes known as the {\it parafermion disk}.  Beginning with the $\sutwo$ metric in Euler angle coordinates~\eqref{eq:su2-param}
\be
ds^2_\su = d\theta^2 + \cos^2\theta\, d\psi^2 + \sin^2\theta\,d\phi^2 
\ee
and gauging (axial) translations in $\psi$, the quotient theory has the metric and dilaton
\be
ds^2_{\rm axial} = d\theta^2 + \tan^2\theta\, d\phi^2 
~~,~~~~
e^{2\Phi} = \frac{\gstrsq}{\cos^2\theta}
~~;
\ee
on the other hand, gauging (vector) translations in $\phi$ results in the geometry
\be
ds^2_{\rm vector} = d\theta^2 + \cot^2\theta\, d\psi^2 
~~,~~~~
e^{2\Phi} = \frac{\gstrsq}{\sin^2\theta}  ~~.
\ee
The model is so named because the effective geometry of the quotient has the topology of a disk, and because after stripping off their $\uone$ dependence the $\sutwo$ currents $J^\pm$ are $\bZ_{k_\su}$ parafermions whose chiral algebra controls the structure of correlation functions. 
These two gaugings are related by T-duality, which flips the sign of the left Euler angle $\psi-\phi$ keeping the right Euler angle $\psi+\phi$ fixed, thus interchanging the roles of $\phi$ and $\psi$.  This structure is clear in the metric, in that T-duality inverts the metric coefficient for the circle being dualized.  The geometry is that of a disk, naively with a diverging size of the angular circle at the boundary of the disk; but this divergence is something of a fake from the point of view of string theory.  String theory on the $\sutwo$ group manifold starting point is completely regular, and gauging simply selects the subset of observables that are invariant under the gauge motion, and thus remains completely regular.  The background so obtained is thus not a regular geometry in any fixed T-duality frame; it is perhaps usefully described as a {\it T-fold}
\cite{Dabholkar:2002sy,Hellerman:2002ax,Kachru:2002sk,Hull:2004in},
a structure described in patches much as for the atlas of charts describing a manifold, but where the maps between coordinates in patch overlaps can involve T-duality transformations.  Thus $\frac{\sutwo}{\uone}$ can be thought of as a ``T-fold two-sphere", with northern and southern hemisphere patches related by  T-duality on the aximuthal circle.

For $\sutwo\times\sutwotil$, one canonical choice for the action of $\uone_s$ is a symmetrical embedding
\be
\big( g_\su,g_\sutil \big) \longrightarrow \big( hg_\su h^{-1}, h^{-1} g_\sutil h \big)
\ee
which leads to a T-fold with effective geometry on the quotient
\begin{align}
ds^2 &= d\theta^2+d\thetat^2+\frac1\Sigma\Big( k_\su(k_\su\tight+k_\sutil\sin^2\!\thetat)\cos^2\!\theta\,d\psi^2 + k_\sutil(k_\sutil\tight+k_\su\sin^2\!\theta)\cos^2\!\thetat\,d\psit^2
\nn\\[-.1cm]
&\hskip 4cm
-k_\su k_\sutil\cos^2\!\theta\cos^2\!\thetat\, d\psi\,d\psit + 4k_\su k_\sutil\sin^2\!\theta\sin^2\!\thetat\, d\chi^2 \Big)
\\[.2cm]
B &= \frac{2k_\su k_\sutil}\Sigma\Big( \cos^2\!\theta \sin^2\!\thetat\, d\chi\wedge d\psi + \sin^2\!\theta \cos^2\!\thetat\,d\chi\wedge d\psit \Big)
\\[.2cm]
e^{-2\Phi} &= \frac{1}{g_s^2} \Sigma
\\[.2cm]
\Sigma &= k_\su\sin^2\!\theta + k_\sutil\sin^2\!\thetat 
\end{align}
since the isometry being gauged has fixed points at $\theta=\thetat=0$ which are flipped to $\theta=\thetat=\frac\pi2$ under simultaneous T-duality (\ie\ flipping from vector to axial gauging) on both groups.

With this angular geometry in hand, we can then apply null gauging, using the $\uone_L\times\uone_R$ null isometries
\be
\big( g_\sl,g_\su,g_\sutil \big) \longrightarrow \big(  h_\ell\,g_\sl \, h_r,  h_\ell^\alpha g_\su  h_r^{\alpha},  h_\ell^{\beta}g_\sutil  h_r^{\beta} \big)
~,~~~
\alpha=\frac{-k_\sutil}{k_\su+k_\sutil}
~,~~~
\beta=\frac{-k_\su}{k_\su+k_\sutil}
\ee
where $\alpha,\beta$ are chosen so that the left and right null Killing vectors $\xi_\ell,\xi_r$ associated to these isometries are null and orthogonal to the Killing vector $\xi_s$ associated to $\uone_s$.  These choices guarantee that the gauging is anomaly free.   The result will be a linear dilaton geometry whose asymptotic angular geometry is the above T-fold, with a fivebrane source singularity in the effective geometry at the origin $\rho=\theta=\thetat=0$ in the above patch, and at $\rho=0,\theta=\thetat=\frac\pi2$ in the T-dual patch.

As in section~\ref{sec:ArgyresDouglas}, worldsheet dualities provide an alternative proposed interpretation for this background~\cite{Giveon:1999zm,Giveon:1999px}.  The orbifold of the tensor product
\be
\bigg(\frac{\sltwo}{\uone}\times \frac{\sutwo}{\uone}\times \frac{\sutwotil}{\uone}\bigg)\Big/\bZ_{n} ~,
\ee
where $n=\textit{lcm}(k_\su,k_\sutil)$, has the same spectrum as the null gauged model above.  The Landau-Ginsburg interpretation of the coset orbifold is that the $\Neqtwo$ superpotential $\cW(\sfZ,\widetilde\sfZ)$ describes a slightly resolved Calabi-Yau hypersurface singularity
\be
\sfZ^{k_\su} + \widetilde\sfZ^{k_\sutil}+\sfZ_1^2+\sfZ_2^2 = \mu
\ee
in $\bC^4$.  The fivebrane interpretation of the background is mirror to this hypersurface singularity, however in this case the angular space of the fivebrane throat is not a conventional geometry but rather a T-fold.

It is not clear that an alternate interpretation is available that would relate the zeroes of the superpotential to the underlying fivebrane locus.

%%%%%%%%%%%%%%%%%%%%%%%%%%%%%%%%%%%%%%%%%%%
%%%%%%%%%%%%%%%%%%%%%%%%%%%%%%%%%%%%%%%%%%%
\subsection{Option 2: Asymmetric gauging and $T^{p,q}$ angular geometries}
\label{sec:angular geom}

Another choice of spatial gauging involves an asymmetric embedding known as the GMM coset of $\sutwo\times\sutwotil$~\cite{Guadagnini:1987ty}, is defined by the $\uone$ isometry
\be
\label{GMMisometry}
\big( g,\gtil \big) \longrightarrow \big( gh^p , h^{q}\gtil  \big)  ~~,
\ee
which we now discuss on its own before incorporating it as the angular geometry of a fivebrane throat.
There are no fixed points of this isometry, and therefore the gauged theory will have an effective geometry free of singularities, which resembles the geometry of the Sasaki-Einstein manifold $T^{p,q}$~\cite{Candelas:1989js} and so we will call it by that name.

%%%%%%%%%%%%%%%%%%%%

Anomaly cancellation imposes a relation between the levels $\nf,\nft$ of the two $\sutwo$'s
\be
\label{anomaly free levels}
\nf=q^2 n
~~,~~~~
\nft=p^2 n~.
\ee
The gauge orbit is generated by the Killing vector
\be
\xi_1=p\,\partial_\phir+q\,\partial_{\philt}
\ee
on $\sutwo\times\sutwotil$, where $\phil,\phir$ are the left/right Euler angles
\be
\phil = \hf(\psi-\phi)
~~,~~~~
\phir = \hf(\psi+\phi)
\ee 
and similarly for $\philt,\phirt$.  
With this choice of gauging, we can choose to parametrize 
\eq{
\phir=p(\gamma+\chi)\quad, \quad \philt=q(\gamma-\chi)~,
}
where $\gamma$ is the gauge parameter and $\chi$ is the remaining physical degree of freedom. Let us pick conventions where $\phir,\philt\in[0,2\pi)$. Since we are gauging a $U(1)$ group and $p,q$ are coprime, the closure of the gauge orbit implies that $\gamma\in \big[0,2\pi\big)$. 

Note that the period of $\chi$ is smaller than $2\pi$ due to gauge invariance.  Consider the shift
\eq{
\chi\longmapsto \chi+2\pi\left(\frac{r_1}{2p}+\frac{r_2}{2q}\right)~,~~~r_1,r_2\in\bZ ~,
}
which leads to 
\eq{
\delta_\chi\phir = 2\pi\left(\frac{r_1}{2}+\frac{r_2p}{2q}\right)
\quad , \quad 
\delta_\chi\philt = -2\pi\left(\frac{r_1 q}{2p}+\frac{r_2}{2}\right)~. 
}
Now consider the compensating gauge transformation 
\eq{\label{alphatrans}
\gamma \longmapsto \gamma +2\pi\left(\frac{r_1}{2p}-\frac{r_2}{2q}\right)~,
}
which leads to
\eq{
\delta_\gamma \phir=2\pi\left(\frac{r_1}{2}-\frac{r_2p}{2q}\right)
\quad, \quad 
\delta_\gamma \philt=2\pi\left(\frac{r_1q}{2p}-\frac{r_2}{2}\right) ~.
}
The combined transformation shifts $\varphi_r,\tilde\varphi_l$ by an integral number of periods.  
Since $p,q$ are coprime, ${\rm min}_{r_1,r_2}\{p r_2+qr_1\in \IZ_+\}=1$. Thus, we can conclude that the range of the $\chi$ coordinate is 
\eq{
\chi\in \big[0,\coeff{2\pi }{2pq}\big)~.
}
We can then rescale to a coordinate $\chihat$ with canonical periodicity $\chihat\in[0,2\pi)$ via
\eq{
\chihat = 2pq\,\chi ~.
}

%%%%%%%%%%%%%%%%%%%%%%%%%%%%%%

Gauging the isometry~\eqref{GMMisometry} leads to the effective metric on the quotient~\cite{PandoZayas:2000he}
\begin{align}
\label{Tpq metric}
ds^2 &=
k_\su \Bigl(d\theta^2+\sin^2 2\theta\,d\phil^2\Bigr) + k_\sutil \Bigl(d\thetat^2+\sin^2 2\thetat\,d\phirt^2\Bigr)
\nn\\
&\hskip1cm
+\frac{\sqrt{k_\su k_\sutil}}{pq}\Bigl(  d\chihat + q\cos2\theta\,d\phil - p\cos2\thetat\,d\phirt \Bigr)^2
\end{align}
which is a circle fibered over $S^2\times S^2$ (or a squashed $S^3$ fibered over $S^2$, in two different ways).%
\footnote{Note that because we have chosen conventions where $\phir,\philt\in[0,2\pi)$, the angles $\phirt,\phil\in[0,4\pi)$.  As a result, the two $\bS^2$'s have conical defects at their poles.  These are not standard orbifold singularities, and don't seem to be a problem for string theory.
}  
This geometry suggests that we define the fiber one-form
\be
\label{gmm fiber 1form}
\mub = \Big( d\chihat + q\cos2\theta\,d\phil - p\cos2\thetat\,d\phirt\Big)
\ee
which allows us to write the metric as 
\eq{
ds^2 &=
k_\su \Bigl(d\theta^2+\sin^2 2\theta\,d\phil^2\Bigr) + k_\sutil \Bigl(d\thetat^2+\sin^2 2\thetat\,d\phirt^2\Bigr)
+\frac{\sqrt{k_\su k_\sutil}}{pq} \mub^2  ~.
}
One also has the effective B-field 
\be
\label{Basymp}
B =  \frac{k_\su}{q} \cos2\theta\,\mub\wedge d\phil + \frac{k_\sutil}{p} \cos2\thetat \,\mub\wedge d\phirt 
+ \sqrt{k_\su k_\sutil} \cos2\theta\cos2\thetat \,d\phil\wedge d\phirt
\ee 
so that the three form flux reduces to a sum of the volume forms on the angular three-cycles one can build out of either of the two $S^2$'s together with the $\chi$-circle fiber (ignoring the last term whose H-flux does not thread a homology cycle).  
We see that there are respectively $2\frac{k_\su}{q}=2n q$ and $2\frac{k_\sutil}{p}=2n p$ units of $H$ flux on the two angular three-cycles.

There are also KK monopoles in the geometry, as one sees from the structure of the $\chi$ circle fibration.  This feature can also be seen by performing a T-duality on the $\chi$ circle, that turns these KK monopoles into NS5-brane fluxes, and the NS5-branes into KK monopoles (see below).  Using the standard Buscher rules~\cite{Buscher:1987qj}
\begin{align}
\Gtil_{\chihat\chihat} &= 1/{G_{\chihat\chihat}}
~~,~~~~
\Gtil_{\chihat\phil} = {B_{\chihat\phil}}/{G_{\chihat\chihat}}
~~,~~~~
\Gtil_{\chi\phirt} = {B_{\chihat\phirt}}/{G_{\chihat\chihat}}
\nn\\[.2cm]
\Gtil_{\phil\phil} &= G_{\phil\phil}-{\big(G_{\chihat\phil}^2-B_{\chihat\phil}^2\big)}/{G_{\chihat\chihat}}
~~,~~~~
\Gtil_{\phirt\phirt} = G_{\phirt\phirt}-{\big(G_{\chihat\phirt}^2-B_{\chihat\phirt}^2\big)}/{G_{\chihat\chihat}}
\nn\\[.2cm]
\Gtil_{\phil\phirt} &= G_{\phil\phirt}-{\big(G_{\chihat\phil}G_{\chihat\phirt}-B_{\chihat\phil}B_{\chihat\phirt}\big)}/{G_{\chihat\chihat}}
\\[.2cm]
\Btil_{\chihat\phil} &= G_{\chihat\phil}/G_{\chihat\chihat}
~~,~~~~
\Btil_{\chihat\phirt} =  G_{\chihat\phirt}/G_{\chihat\chihat}
\nn\\[.2cm]
\Btil_{\phil\phirt} &= B_{\phil\phirt} - \big(G_{\chihat\phirt}B_{\chihat\phil} - B_{\chihat\phirt} G_{\chihat\phil}\big)/G_{\chihat\chihat}
\nn
\end{align}
one finds in the T-dual frame
\be
\Btil_{\chihat \phirt} \sim  -p \cos2\thetat\,d\mub\wedge d\phirt
~~,~~~~
\Btil_{\chihat \phil} \sim  q \cos2\theta\,d\mub\wedge d\phil  ~,
\ee
indicating that there are KKM charges $2 p$ and $2 q$ along the corresponding cycles in the original duality frame, consistent with the coefficient of the connection in the $\mub$ fiber which codes the KKM charge.

%%%%%%%%%%%%%%%%%%%%%%%%%%%%%%%%%%%%%%%%%%%
%%%%%%%%%%%%%%%%%%%%%%%%%%%%%%%%%%%%%%%%%%%
\subsection{Combining squashed $T^{p,q}$ and null gauging}

We can now pair a squashed version of this angular space with null gauging to generate a background with intersecting fivebranes and KK monopoles.   Consider squashing both $\sutwo$ and $\widetilde{SU}(2)$, and as above gauge the spatial isometry acting on the left on one factor and on the right on the other factor
\be
\label{GMM KV}
\xi_1= \tilde\alpha_l\,\partial_\philt+\alpha_r \,\partial_{\phir} \equiv \zeta_L^{~} + \zeta_R ^{~} ~;
\ee
here $\phil,\phir$ are the left/right Euler angles of the squashed three-sphere
\be
\phil = \half\Big(R_\su\psi-\frac1{R_\su}\phi\Big)
~~,~~~~
\phir = \half\Big(R_\su\psi+\frac1{R_\su}\phi\Big)
\ee 
and similarly for $\philt,\phirt$.  In addition we gauge left and right null vectors orthogonal to $\xi_1$
\begin{align}
\label{asym null KVs}
\xi_2&= \partial_\tau + \partial_\sigma  - \tilde\beta_r\,\partial_{\phirt}
\nn\\
\xi_3&= \partial_\tau - \partial_\sigma - \beta_l\,\partial_\phil 
~~.
\end{align}
Anomaly cancellation imposes
\be
c_s \alpha_r = \beta_l = \sqrt{\frac{\nz}{\nf}}  =  \sqrt{\frac{\nft}{\nf+\nft}}
~~,~~~~
c_s \tilde\alpha_l = \tilde\beta_r = \sqrt{\frac{\nz}{\nft}}  =  \sqrt{\frac{\nf}{\nf+\nft}}
\ee
where we have used~\eqref{n0 value}; $c_s$ is an arbitrary constant.  Commensurability of the periodicities of the Euler angles and the $\uone$ gauge orbits leads us to parametrize the levels as
\be
\nf = q^2n
~~,~~~~
\nft = p^2n
~~,~~~~
\nz = \frac{p^2q^2n}{p^2+q^2}
\ee
and we can use the arbitrary constant $c_s$ to set $\tilde\alpha_l=q,\alpha_r=p$.

The contribution to the action from gauging is
\eq{
\CS_{gauge} &= \int \Big[-2A^2\bar{A}^3(-\xi^i_2 G_{ij}\xi^j_3)
-2 A^2(\xi_2^i G_{ij}\bar\partial\varphi^j)-2\bar{A}^3 (\xi^i_3 G_{ij}\partial \varphi^j)
\\[.1cm]
&\hskip 1cm
-A^1 \bar{A}^1(-\xi^i_1G_{ij}\xi_1^j)-2 A^1(\xi^i_1 G_{ij}\bar\partial \varphi^j)-2\bar{A}^1(\xi^i_1 G_{ij}\partial \varphi^j)
\\[.2cm]
&\hskip 2cm
-2 A^1 \bar{A}^3(-\xi^i_1 G_{ij}\xi^j_3)-2A^2\bar{A}^1(-\xi^i_2G_{ij}\xi^j_1)
\Big]~.
}
Here, the first line gives the contribution of the null isometries $\xi_2,\xi_3$; the second line results from gauging the spatial isometry $\xi_1$, and the third line represents an interaction between the two.
Thus, we find that we can write the part of the action that couples to the gauge field as 
\be
\cS_{\rm gauge} = -\int \big(  2\CA \bar{\J}+\text{2}\bar{\cA}\J +\text{4}{\cA}^{\!\it T} \M \, \bar\cA \big)
\ee
where $\cA=(A^1,A^2)$, $\bar\cA=(\bar A^1,\bar A^3)$, 
$\J=(J_{\sst GMM},\cJ)$, $\bar \J = ( \bar J_{\sst GMM},\bar\cJ)$
and
\eq{
\M&= \frac{k_{sl}}{2} 
\left(\begin{array}{cc}
1 & \ff(\theta,R_\su)  \\[.1cm]
- \ff(\thetat,R_\sutil) & ~ \ff(i\rho,R_\sl)
\end{array}\right)
~,\\
}
where $f(x,R)$ was defined in~\eqref{fdefs}.
Upon integrating out the gauge fields, we find that the resulting contribution to the action is given by 
\eq{
\cS_{\rm gauge}= \int \J^T \M^{-1} \bar{\J}~.
}
The effective geometry will have apparent singularities at the vanishings of the determinant of $M$:
\begin{align}
\Sigma = {\det}[M] 
= \nz^{\;2} \Big[ \ff(i\rho,R_\sl) - \ff(\theta,R_\su)\fft(\thetat,R_\sutil)\Big] 
\end{align}
which occur at $\rho\tight=0,\theta\tight=0,\thetat\tight=\frac\pi2$ and $\rho\tight=0,\theta\tight=\frac\pi2,\thetat\tight=0$,
though of course the singularity is an artifact of the supergravity approximation; the full string theory is completely smooth so long as the fivebranes are completely separated on their Coulomb branch.

Now we can pick a gauge in which $\tau\tight=\sigma\tight=0$ for gauging the null currents; for the asymmetric gauging of the GMM currents that yields $T^{p,q}$ for the angular space, one can choose a parametrization 
\be
\philt = q\, \big(\gamma-\chi\big)
~~,~~~~
\phir = p\,\big(\gamma+\chi\big)
\ee 
in terms of a coordinate $\gamma$ along the gauge orbit and a physical coordinate $\chi$ that parametrizes the source circle.

Setting the gauge parameter $\gamma=0$ and integrating out the gauge fields, one finds an effective geometry
\begin{align}
\label{GMM Geff}
ds^2 &=
\frac{np^2q^2}{p^2+q^2} \,d\rho^2 + nq^2 \,d\theta^2 +np^2 \,d\thetat^2
\nn\\[.1cm]
&\hskip .5cm
+ \frac{1}{\ff_+^\su \ff_+^\sutil \ff_-^\sl + \ff_-^\su \ff_-^\sutil \ff_+^\sl} \Big[
\big(\ff_+^\su \ff_+^\sutil \ff_-^\sl - \ff_-^\su \ff_-^\sutil \ff_+^\sl \big) \big(4np^2q^2\,d\chi^2 + nq^2\, d\phil^2 + np^2\, d\phirt^2\big)
\nn\\[-.1cm]
&\hskip 5cm
+npq \big(\ff_+^\su\ff_+^\sutil\ff_+^\sl - \ff_-^\su\ff_-^\sutil\ff_-^\sl \big) \,d\phil\,d\phirt
\nn\\[.1cm]
&\hskip 5cm
+2npq^2\big(\ff_-^\su\ff_+^\sutil\ff_-^\sl - \ff_+^\su\ff_-^\sutil\ff_+^\sl \big) \,d\chi\, d\phil
\\[.1cm]
&\hskip 5cm
-2nqp^2\big(\ff_+^\su\ff_-^\sutil\ff_-^\sl - \ff_-^\su\ff_+^\sutil\ff_+^\sl \big) \,d\chi\, d\phirt
\Big]
\nn\\[.4cm]
\label{GMM Beff}
B &= \frac{1}{\ff_+^\su \ff_+^\sutil \ff_-^\sl + \ff_-^\su \ff_-^\sutil \ff_+^\sl} \Big[
npq\, \big( \ff_+^\su\ff_+^\sutil\ff_+^\sl + \ff_-^\su\ff_-^\sutil\ff_-^\sl \big)\,d\phil\wedge d\phirt
\nn\\[-.1cm]
&\hskip 4.5cm
2npq^2\, \big( \ff_+^\su\ff_-^\sutil\ff_+^\sl + \ff_-^\su\ff_+^\sutil\ff_-^\sl \big)\,d\chi\wedge d\phil
\\[.1cm]
&\hskip 4.5cm
2nq^2p\, \big( \ff_+^\sutil\ff_-^\su\ff_+^\sl + \ff_-^\sutil\ff_+^\su\ff_-^\sl \big)\,d\chi\wedge d\phirt \Big] ~.
\nn\\[.2cm]
\label{GMM Phieff}
e^{-2\Phi} &= \frac{1}{g_s^2}\Big({\ff_+^\su \ff_+^\sutil \ff_-^\sl + \ff_-^\su \ff_-^\sutil \ff_+^\sl}\Big)
\end{align}
This geometry combines the linear dilaton structure of fivebrane throats with the squashed $T^{p,q}$ angular geometry.  To understand its structure, we now consider various limits, which will lead us to a physical interpretation of this background.

%%%%%%%%%%%%%%%%%%%%%%%%%%%%%%%%%%%%%%%%%%%
%%%%%%%%%%%%%%%%%%%%%%%%%%%%%%%%%%%%%%%%%%%
\subsection{Asymptotic geometry}

The geometry simplifies considerably in the asymptotic limit $\rho\to\infty$.  We further choose $R_\sl=1$, which simplifies many expressions; for instance, 
\be
\ff_+(i\infty,R_\sl\tight=1)=1
~~,~~~~
\ff_-(i\infty,R_\sl\tight=1)=\infty  ~;
\ee
furthermore, the angular geometry will have the form of a squashed $T^{p,q}$ space, and the spectrum matches a coset orbifold construction, only for this choice of $R_\sl$.  
It is useful to define an asymptotic fiber one-form as in~\eqref{gmm fiber 1form}
\be
\mub =  d\chihat + q\,\ff(\theta,R_\su)\, d\phil
- p\,\ff(\thetat,R_\sutil)\, d\phirt
\ee
(where $f(x,R)$ was defined in~\eqref{fdefs}),
so that the asymptotic metric takes the form
\begin{align}
ds^2 &\sim \frac{np^2q^2}{p^2+q^2}\, d\rho^2 + n\,d\mub^2 
\\[.2cm]
&\hskip1cm
+ nq^2 \Big[ d\theta^2 + \Big(\coeff{\ff_+^2(\theta)-\ff_-^2(\theta)}{\ff_+^2(\theta)}\Big)\,d\phil^2 \Big]
+ np^2 \Big[ d\thetat^2 + \Big(\coeff{\ff_+^2(\thetat)-\ff_-^2(\thetat)}{\ff_+^2(\thetat)}\Big)\,d\phirt^2 \Big] ~.
\nn
\end{align}
From the form~\eqref{fdefs} of the functions $f_\pm$, one has
\be
\frac{\ff_+^2(\theta)-\ff_-^2(\theta)}{\ff_+^2(\theta)} = \frac{ R_\su^2 \sin^2 2\theta}{(\cos^2\theta+R_\su^2\sin^2\theta)^2}
\ee
and similarly for $\ff_\pm(\thetat)$; the asymptotic angular geometry is roughly a circle double fibered over two squashed two-spheres (note however that there is a conical defect at the poles of the $\bS^2$'s), \ie\ it is a sort of squashed version of $T^{p,q}$. 

The asymptotic $B$-field has the form
\begin{align}
\label{R31 Basymp}
B &= nq\Big(\frac{ \cos^2\theta-R_\su^2\sin^2\theta}{\cos^2\theta+R_\su^2\sin^2\theta}\Big) \,d\hat \chi\wedge d\phil
+ np\Big(\frac{ \cos^2\thetat-R_\sutil^2\sin^2\thetat}{\cos^2\thetat+R_\sutil^2\sin^2\thetat}\Big) \,d\hat \chi\wedge d\phirt
\nn\\[.2cm]
&\hskip 1cm
+ npq\Big(\frac{ \cos^2\theta-R_\su^2\sin^2\theta}{\cos^2\theta+R_\su^2\sin^2\theta}\Big)
\Big(\frac{ \cos^2\thetat-R_\sutil^2\sin^2\thetat}{\cos^2\theta+R_\sutil^2\sin^2\thetat}\Big)\,d\phil\wedge d\phirt ~.
\end{align}
%\
Thus there are $2nq$ and $2np$ fivebranes transverse to the corresponding 3-cycles~-- the squashing of the asymptotic angular geometry doesn't change the quantization conditions on the $H$ flux.  The flux on each cycle is split into two parts, such that there are two intersection points in~$\theta,\thetat$ of two stacks of fivebranes.

%%%%%%%%%%%%%%%%%%%%%%%%%%%%%%%%%%%%%%%%%%%
%%%%%%%%%%%%%%%%%%%%%%%%%%%%%%%%%%%%%%%%%%%
\subsection{Near-source structure}

The source structure is revealed in the opposite limit $\rho\to 0$.  The metric coefficients in the angular space parametrized by $(\chi,\phil,\phirt)$ are all proportional to
\be
\frac{R_\sutil^2\cos^2\!\theta \sin^2\!\thetat \pm R_\su^2 \sin^2\!\theta \cos^2\!\thetat}{\cos^2\!\theta \cos^2\!\thetat + R_\su^2 R_\sutil^2 \sin^2\!\theta \sin^2\!\thetat} ~,
\ee
and one can check that the circle parametrized by $q\phil+p\phirt$ degenerates in this limit.  In addition, the proper size of the $\chi$ circle blows up at the corners of the square parametrized by $\theta,\thetat$, specifically at $\theta=0,\thetat=\frac\pi2$ and $\theta=\frac\pi2,\thetat=0$.  At the opposite corners of the $\theta,\thetat$ square, namely at $\theta=\thetat=0$ and at $\theta=\thetat=\frac\pi2$, all the circles parametrized by $\chi,\phil,\phirt$ degenerate, and one finds an orbifold singularity of multiple KK monopoles.  The roles of these diagonally opposite corners of the square parametrized by $\theta,\thetat$ are interchanged by T-duality on the $\chi$ circle, which interchanges the roles of NS5-branes and KK monopoles.

%%%%%%%%%%%%%%%%%%%%%%%%%%%%%%%%%%%%%%%%%%%
%%%%%%%%%%%%%%%%%%%%%%%%%%%%%%%%%%%%%%%%%%%
\subsection{Physical interpretation of the asymmetrically gauged model}
\label{sec:phys interp}

Using the now well-understood construction of fivebranes on the Coulomb branch using null gauging, we can piece together a heuristic picture of the geometry~\eqref{GMM Geff}-\eqref{GMM Phieff} and its fivebrane interpretation.  A single stack of $\nfive$ coincident fivebranes sources the geometry
\be
ds^2 = ds^2_{\bR^{5,1}} + H\big(dr^2 + r^2 d\Omega_3^2 \big)  ~~,~~~~ H=1+\frac{\nfive\lstr^2}{r^2} ~.
\ee
As one approaches the fivebrane source, the size of the angular $S^3$ shrinks, and then stabilizes at the fivebrane charge radius $r_5 = \sqrt{\nfive}\,\lstr$; the longitudinal directions have no warping.  Coming toward the intersection locus of two stacks of fivebranes oriented as
\begin{align}
12\mathsf{xxxx}789\nn\\
\mathsf{xx}34\mathsf{xx}789
\end{align}
the directions $12$ and $34$ are transverse to only one stack, and so are stabilized by the above effect.  On the other hand, the $56$ direction is transverse to both stacks, and so will pick up two copies of the warp factor $H$, one from each stack; one expects the angular circle in the $56$ direction (the circle we have parametrized by $\chi$) to blow up at the source, but not the angular directions 12 and 34 that are transverse to one stack but not the other.  Even though the metric looks singular at the places where the $\chi$ circle blows up, the exact worldsheet string theory is nonsingular, and so one expects (as was the case for the single stack) that the fivebranes are spread out along this circle, and that this spreading of the branes onto their Coulomb branch is responsible for blocking strings from reaching a strong coupling region near the fivebranes.

However, in the decoupled fivebrane throat described by the geometry~\eqref{GMM Geff}-\eqref{GMM Beff} we have two intersection points, at $\rho=0$ together with $\theta=\frac\pi2\tight-\thetat=0,\frac\pi2$.  The $H$ flux associated to~\eqref{R31 Basymp} indicates that we have $n p$ and $n q$ branes intersecting at each of these points; the branes are spread out along the $\chi$ circle transverse to both branes, which loosely speaking descends from the angular direction in the 56 plane.  Similarly, one has KK monopoles of total charge $ p$ and $ q$ intersecting at $\rho=0$ and $\theta=\thetat=0,\frac\pi2$, also spread out along the $\chihat$ circle (or equivalently, NS5-branes spread out along the T-dual $\tilde\chi$ circle).
The worldsheet description of the throat of the intersection of a single stack of NS5-branes with a single stack of KK monopoles was worked out in~\cite{Kutasov:1998zh}, involving an $\sutwo$ WZW model at level $\nfive n_{\rm KK}$.  The null-gauged model above in a sense generalizes this construction to the intersection of two stacks of fivebranes and two stacks of KK monopoles (with each stack of each type split into two bunches, one at either pole of the angular $\IS^2$'s).  The angular geometry can be thought of topologically as $\IS^3\times\IS^2$ in two different ways.  The underlying pair of WZW models have levels $np^2$ and $nq^2$, and describe $np$ NS5's and $p$ KKM's each in the pair of stacks associated to one three-sphere, and $nq$ NS5's and $q$ KKM's each in the pair of stacks associated to the other three-sphere.

The T-duality of section~\ref{sec:angular geom} is a straightforward consequence of null gauging.  We simply flip the sign of $J_{\sst GMM}$, which interchanges the roles of $\gamma$ and $\chi$; the fibration of the $\chi$ circle and the B-fluxes involving this circle are interchanged.  Now the fivebrane singularities are at $\theta=\thetat=0,\frac\pi2$ with the KKM intersections at $\theta=\frac\pi2-\thetat=0,\frac\pi2$).

In the analysis of~\cite{Dasgupta:1998su}, a chain of duality transformations was employed to relate D3-branes at an orbifold of the conifold to NS5-branes intersecting over $\bR^{3,1}$.  At some point in their analysis it was necessary to ``compactify'' the 12 and 34 planes; we see here that this procedure is not so innocent, as the ``point at infinity'' that is being added is the locus of another fivebrane singularity.

%%%%%%%%%%%%%%%%%%%%%%%%%%%%%%%%%%%%%%%
%%%%%%%%%%%%%%%%%%%%%%%%%%%%%%%%%%%%%%%

\subsection{Spectrum}

The spectrum of the GMM model follows from the same sort of considerations that arose in the Argyres-Douglas models.  Because the asymptotic angular geometry only approached the (squashed) $T^{p,q}$ form for $R_\sl=1$, we will need to squash the two $\sutwo$ factors in order to have nontrivial solutions to the gauge constraints.  We set the $\sutwo$ squashing parameters
\be
R_\su = R_\sutil = \frac{\sqrt{p^2+q^2}}{p+q} ~.
\ee
The analysis of the spectrum then follows along the same lines.  In particular, the BPS spectrum consists of highest weight states of the various (squashed) groups, with the quantum numbers related via
\begin{align}
\label{gmm BPS spectrum}
\jsu = \msu &= \bmsu = q r
~~,~~~~
\jsut = \msut = \bmsut = p r
~~,~~~~
\jsl = \msl = \bmsl = 1 + \frac{rpq(p+q)}{p^2+q^2}
\nn\\[.1cm]
r &= 0,\hf,\ldots,\textrm{min}\Big(\frac{qn}2-1,\frac{pn}2-1,\frac{np^2q^2-p^2-q^2}{2pq(p+q)} \Big)
~.
\end{align}
The shift by one in $\jsl$ is compensated by an oscillator excitation of $\psi_\sl^-$ on both left and right that carries the supergraviton polarization of the state (in the $-1$ picture for the superghosts).  
Once again there are additional choices where the polarization is taken in the $\sutwo$ factors; and there are analogous BPS RR sector vertices.
For these choices, the null gauge constraints as well as the spatial GMM gauge constraints
\eq{
J^3_\sl = \sqrt{\frac{k_\sl}{k_\su}} \,\Jsq_\su
~~,~~~~
\bar J^3_\sl = \sqrt{\frac{k_\sl}{k_\sutil}} \,\Jbsq_\sutil
~~,~~~~
p\Jsq_\sutil = q\Jbsq_\su
}
are satisfied, as are the Virasoro constraints analogous to~\eqref{AD Vir} (note though that the contributions from $p_L^{\sutil}$ and $p_R^\su$ are removed from these constraints by the spatial GMM gauging).  Again, there are other possibilities where one takes the fermion polarization in $\sutwo,\sutwotil$.  And again, these supergraviton states have ``winding tachyon'' equivalents via generalized FZZ duality.  One can check that the spectrum~\eqref{gmm BPS spectrum} matches what one obtains from a coset orbifold construction using GMM gauging on $\sutwo\times\sutwotil$.  

%%%%%%%%%%%%%%%%%%%%%%%%%%%%%%%%%%%%%%%%%%%
\subsection{Triple intersections}

It seems clear that the considerations of this section apply more generally, for instance one can add another $\sutwo$ factor to $\cG$, and another $\uone$ factor to $\cH$.  In this way one should be able to describe intersections of three stacks of fivbranes.  We leave the details to the interested reader.

%%%%%%%%%%%%%%%%%%%%%%%%%%%%%%%%%%%%%%%
%%%%%%%%%%%%%%%%%%%%%%%%%%%%%%%%%%%%%%%

\section{Supertubes}
\label{sec:supertubes}

The null gauging framework reveals a close connection between Coulomb branch fivebranes and the two-charge BPS states known as supertubes.  The basic idea is to modify the null isometries to add an admixture of a null isometry tangent to the fivebrane worldvolume.  Then the structure $\cJ\bar\cJ/\Sigma$ in the effective geometry that arises upon elimination of the gauge fields incorporates terms involving the square of the longitudinal currents, which encodes either momentum or fundamental string charge carried by the background (depending on a choice of signs in the null currents); in addition one has a cross-term between the longitudinal and transverse components of the null currents, which encodes angular momentum carried by the background.  As noted in~\cite{Martinec:2017ztd}, this procedure is completely general, so we can apply it to any of the Coulomb branch fivebrane configurations described above.

In the worldsheet formalism of gauged WZW models describing fivebrane backgrounds, Coulomb branch fivebrane configurations are related to NS5-P and NS5-F1 supertubes by a suitable modification of the null isometries being gauged:
\eq{
\label{supertube currents}
\cJ_{\textrm{supertube}} = \cJ_{\textrm{Coul}} - \alphab \big( \partial t -\vareps \partial y \big)
~~,~~~~
\bar \cJ_{\textrm{supertube}} = \bar \cJ_{\textrm{Coul}} - \alphab \big( \bar\partial t - \bar \partial y \big)
}
where $\vareps=1$ for NS5-P supertubes and $\vareps=-1$ for NS5-F1.

Consider for example the Argyres-Douglas background of section~\ref{sec:ArgyresDouglas}, which describe $k_\su=\nfive$ NS5-branes wrapped on a slightly resolved conifold (or its fourfold analogue, in the $\bR^{1,1}$ analogue of section~\ref{sec:2d version}), and spread in a circular array on their Coulomb branch.  We parametrize the fivebrane worldvolume by $(x,y,z,t)$ and compactify the spatial directions on $\bS^1_y\times \bT^2_{xz}$ where the $\bT^2$ cycles are of order the string scale and $\bS^1_y$ has radius $\Rytil$; following~\cite{Martinec:2017ztd}, we choose $\alphab=k/\Rytil$ for NS5-P; T-duality flips the sign of $\vareps$ while sending $\Rytil\to1/\Ry$, so that $\alphab=k\Ry$.  For the 2d version, one simply drops the $\bT^2$.

\begin{figure}[h!]
\centering
\includegraphics[width=.4\textwidth]{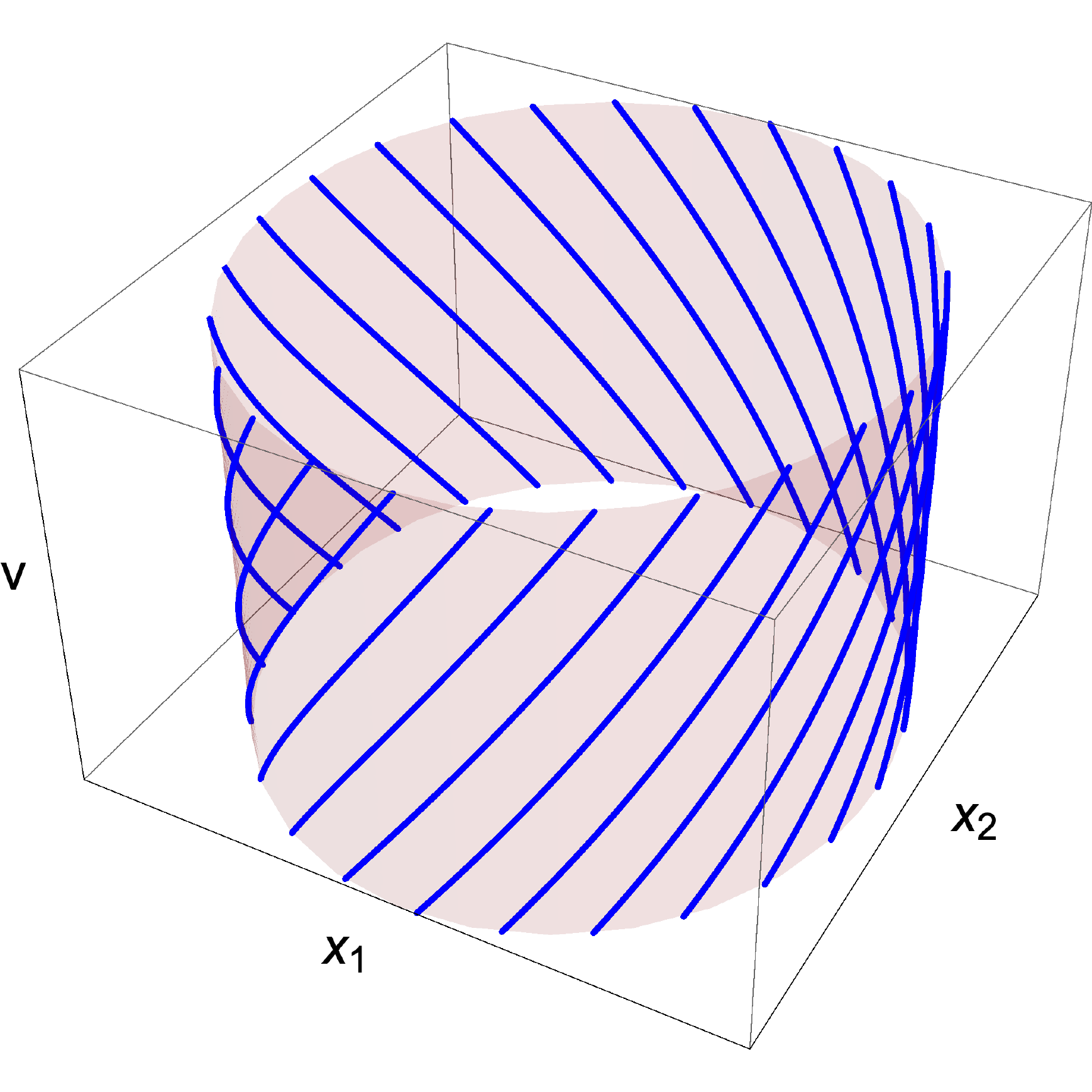}
\caption{\it Source profile for the round NS5-P supertube, with $\nfive=25$ and $k=4$.}
\label{fig:CircularSupertube}
\end{figure}

The interpretation is particularly transparent in the NS5-P frame, where it parallels the structure found for fivebranes wrapped on $\bS^1_y\times \bT^4$ studied in~\cite{Martinec:2017ztd,Martinec:2018nco,Martinec:2019wzw,Martinec:2020gkv}.  The term $\cJ\bar\cJ/\Sigma$ in the worldsheet effective action induces $dt \, dy$ and $dt\, d\psi$ cross terms in the effective metric coefficients that reflect the momentum carried by the fivebranes along the $y$ circle, and their angular momentum along the transverse $\bS^3$.  Several diagnostics indicate that the model describes a situation where the fivebranes have been tilted over to wrap the $(\nfive,k)$ cycle of the angular two-torus $\bT^2_{y\psi}$ (see figure~\ref{fig:CircularSupertube}).  If $k$ and $\nfive$ are mutually prime, there is a single fivebrane that wraps the $y$ circle $\nfive$ times, otherwise there are $\textit{gcd}(k,\nfive)$ interleaved fivebranes.

On the surface singularity side of the Calabi-Yau/Landau-Ginsburg duality, the spiraling fivebranes become a set of spiraling Kaluza-Klein monopole cores, and there is a topological $\bS^2$ whose azimuthal direction is the fibered $\bS^1$ of the KKM and whose polar direction is the interval between adjacent KKM cores.  A slice through the geometry at constant $\psi$ reveals $k$ vanishing cycles.  But as the figure suggests, there is actually only a single $\bS^2$ that spirals around, for instance the $k$ cycles at constant $\psi$ undergo a $\bZ_k$ cyclic shift upon transporting them once around the $\psi$ circle.

The structure for the Argyres-Douglas theories is much the same.  There is again a Coulomb branch for the fivebranes, and one can spread them out along a transverse circle (parametrized by $\psi$ in our conventions above).
%
%%%%%%%%%%%%%%%%%%
\vspace{0.5mm}
\vspace{1mm}
%%%%%%%%%%%%%%%%%%
%

The NS5-F1 frame effective geometry takes the form
\begin{align}
ds^2 &= ds^2_{\IT} + k_\sl \, d\rho^2 + k_\su\, d\theta^2 
\nn\\[.1cm]
&\hskip 1cm
+\frac{1}{\Sigma} \Big[ k_\sl \Big(R\cosh^2\!\rho\,\sin^2\!\theta + \frac1R \sinh^2\!\rho\,\cos^2\!\theta \Big)\big(-dt^2+dy^2\big)
\\[.2cm]
&\hskip 2cm
\frac{k_\su}{R} \big(\alphab^2+k_\sl \sinh^2\!\rho\big) \sin^2\!\theta\, d\phi^2
+ {k_\su}{R} \big(\alphab^2+k_\sl \cosh^2\!\rho\big) \cos^2\!\theta\, d\psi^2
\nn\\[.2cm]
&\hskip 2cm
+\sqrt{k_\sl k_\su}\,R\alphab \Big( \cos^2\theta \, dt\,d\psi - \sin^2\theta \, dy\,d\phi \Big)  \Big]
\nn\\[.3cm]
B &= \frac1\Sigma \Big[
\frac{k_\su}{R}\big(\alphab^2 + k_\sl\sinh^2\!\rho\big)\cos^2\!\theta\,d\phi\wedge d\psi
+ \alphab^2\Big(\frac1R\cos^2\!\theta - R\sin^2\!\theta \Big) dt\wedge dy
\nn\\[.2cm]
&\hskip 1.5cm
\sqrt{k_\sl k_\su}\,\alphab\big( \sin^2\!\theta\,dt\wedge d\phi - \cos^2\!\theta\,dy\wedge d\psi \big)
\Big]
\\[.3cm]
e^{-2\Phi} &= \frac{1}{g_s^2 k_\su} \Sigma
\\[.3cm]
\Sigma &= R\big( \alphab^2+k_\sl \cosh^2\!\rho\big)\sin^2\!\theta + \frac1R\big( \alphab^2+k_\sl\sinh^2\!\rho\big)
\end{align}
where $R^2 = k_\su/k_\sl$,%
\footnote{Among the one-parameter family of squashings considered in section~\ref{sec:ADgeom}, here we choose the one with $R_\sl=1$, because the geometry in the cap region $e^{2\rho}\ll \alphab^2$ will involve an undeformed $AdS_3$ rather than a squashed version.  In section~\ref{sec:ADgeom} it did not matter~-- all choices led to the same effective geometry and vertex operator spectrum.  But here the gauging involves $\IR^t\times \IS^1_y$ and distinguishes these choices.  The choice $R_\sl=1$ is motivated by the fact that for $k=1$ the cap geometry should be the spectral flow of the $AdS_3$ vacuum.}
and $ds^2_{\IT}$ is the metric on the additional torus that the fivebranes are compactifies on, \ie\ a four-torus in the critical dimension, two-torus for the Argyres-Douglas theories, and no torus for the compactification down to 2d.  

The geometries in the latter two noncritical theories is again roughly a squashed version of the one in the critical dimension. 
It rolls over at the scale $e^{2\rho}\sim \alphab^2$ from the linear dilaton geometry of the fivebrane throat at large $\rho$ to a deformed $(AdS_3 \times \IS^3)/\IZ_k$ geometry at small $\rho$.  More precisely, we can isolate the cap geometry by taking a further decoupling limit $\Ry\to\infty$ (\ie\ $\alphab\to\infty$)
\eq{
ds^2_{\rm cap} &= ds^2_{\IT} + k_\sl \Big( d\rho^2  
-\cosh^2\!\rho \,\big(dt/\alphab \big)^2  + \sinh^2\!\rho\,\big(dy/\alphab \big)^2 \Big)
\\[.2cm]
&\hskip 1cm
+ k_\su\Big( d\theta^2 + \frac{\big(d\phi -  dy/\alphab\big)^2\sin^2\!\theta + \big(R\, d\psi+dt/(R\alphab)\big)^2\cos^2\!\theta}{\cos^2\!\theta+R^2\sin^2\!\theta}  \Big)
}
In the critical dimension, where $R=1$, the geometry is a spectrally flowed version of $(AdS_3\times \IS^3)/\IZ_k$~\cite{Lunin:2001fv}, while for the lower-dimensional examples, $R\ne 1$ and the $\IS^3$ metric is squashed as well as spectrally flowed.

A natural extension of the Landau-Ginsburg picture~\cite{Martinec:2020gkv} proposes a modified superpotential
\be
\label{ST spotl}
\cW_0 = \sfZ^{k_\su} e^{ik\sfv} - \lambda_0 \, e^{k_\su\sfX} = \prod_{\ell=1}^{k_\su} \Bigl( \sfZ \,e^{ik\sfv/{k_\su}} - \mu_\ell\, e^{\sfX} \Bigr) 
\ee 
to describe the spiraling source profile of figure~\ref{fig:CircularSupertube} in terms of the spiralling zeroes of the modified superpotential.
The 1/2-BPS vertex operators discussed in section~\ref{sec:ArgyresDouglas} continue to be 1/2-BPS in the supertube context.  Their effect is to perturb the 1/2-BPS supertube to nearby 1/2-BPS configurations.
\be
\cW = \cW_0 + \sum_{n}  \dlam_n \, \lambda_0^{k_\su-n-1} \, \sfZ^n e^{(k_\su-n)\sfX} e^{ink\sfv/k_\su} ~.
\ee
Note, however, that only the range $n>\hf k_\su-1$ correspond to normalizable deformations~\cite{Giveon:1999zm}.
In addition the spectral flows of these vertex operators are distinct physical vertex operators~\cite{Martinec:2020gkv}, and describe more complicated deformations of the source profile  
\be
\cW = \cW_0 + \sum_{n} \sum_{w_y} \dlam_{n,w_y} \, \lambda_0^{k_\su-n-1} \sfZ^n \exp\Bigl[{\bigl(k_\su\tight-n\bigr)\sfX + i\Bigl(\frac{nk+ w_yk_\su}{k_\su}\Bigr)\sfv}\Bigr]  
\ee
with winding on the $y$ circle.
It seems clear that for the Argyres-Douglas theories, there will be a large ensemble of 1/2-BPS supertubes obtained by compactifying the theory on $\IS^1_y\times\IT^2$ and working in superselection sectors of F1 winding $n_1$, which in the $AdS_3$ decoupling limit $\Ry\to\infty$ descend to Ramond ground states of the associated spacetime ${\rm CFT}_2$.

For the 2d analogues, the BPS spectrum is more limited, for instance there are no NS sector 1/2-BPS operators in the unitary range.  But as discussed in~\cite{Martinec:2020gkv}, it is precisely such operators that deform the supertube between different values of the orbifold parameter $k$ by changing the underlying state of the background.  Furthermore, the background with $k=1$ is a large gauge transformation (\ie\ spectral flow) of global $AdS_3$ in the decoupling limit, which is thought to be a non-normalizable state~\cite{Giveon:2005mi}.  All this suggests that the different supertube backgrounds exist, but do not belong to a family of states in the same theory; rather, the deformations that map from one to the other are non-normalizable, and therefore change the theory under discussion to a different one.  For instance, in the $k=1$ model the geometry is naively the spectral flow of global $AdS_3\times \IS^3$, but the $AdS_3$ vacuum is supposedly not part of the spectrum of the spacetime CFT.

In the intersecting fivebrane models of sections~\ref{sec:1d-2d intersection} and~\ref{sec:3d intersection}, the story is much the same.  One can add a contribution to the null currents along $t\pm y$ which spins up the fivebranes around the angular spheres while imparting momentum or F1 winding along $\IS^1_y$.
The effective geometry associated to these models in the NS5-F1 duality frame consists of the linear dilaton throat of fivebranes at large radius, as described in the previous section; at a radial scale set by $k\Ry$, the geometry rolls over to an $AdS_3$ cap with a $\bZ_k$ orbifold singularity at the tip.

We start with the gauge current~\eqref{R21 gauging} and tilt the gauging according to~\eqref{supertube currents}, and choose to squash the $\sutwo$ factors but not $\sltwo$ in order that the cap geometry is $AdS_3$ rather than a squashed version of it.  The asymptotic geometry is linear dilaton time squashed $\sutwo\times\sutwotil$, and the $AdS_3$ limit $\alphab\to\infty$ leads again to a spectral flow of $AdS_3$ times squashed $\sutwo\times\sutwotil$
\eq{
ds^2_{\rm cap} &= ds^2_{\IS^1} + k_\sl \,d\rho^2  
+ (k_\su+k_\sutil)\Big( -\cosh^2\!\rho \,\big(dt/\alphab \big)^2  + \sinh^2\!\rho\,\big(dy/\alphab \big)^2 \Big)
\\[.2cm]
&\hskip 1cm
+ k_\su\Big( d\theta^2 + \frac{\big(d\phi +  R_\su dy/\alphab\big)^2\sin^2\!\theta + \big(R_\su\, d\psi-dt/\alphab \big)^2\cos^2\!\theta}{\cos^2\!\theta+R_\su^2\sin^2\!\theta}  \Big)
\\[.2cm]
&\hskip 1cm
+ k_\sutil\Big( d\thetat^2 + \frac{\big(d\phit +  R_\sutil dy/\alphab\big)^2\sin^2\!\thetat + \big(R_\sutil\, d\psit-dt/\alphab \big)^2\cos^2\!\thetat}{\cos^2\!\thetat+R_\sutil^2\sin^2\!\thetat}  \Big) ~.
}
Curiously, the asymptotic geometry is not quite that of the large $\cN\tight=4$ theory, for which the $\sutwo$ factors would be unsquashed.  It seems that the holographic RG flow from (the very-near horizon limit of) the Coulomb branch of fivebranes of section~\ref{sec:3d intersection} to an $AdS_3$ cap lands us on a deformed geometry which does not have large $\cN\tight=4$ enhanced symmetry, but rather only $\Neqtwo$ supersymmetry.

These backgrounds describe specific BPS states in the fivebrane theory, in particular superselection sectors where the fivebranes carry a macroscopic amount of momentum or fundamental string charge.  In the NS5-P frame, these BPS states are labelled by the Fock space of momentum excitations of the wrapped fivebrane; since T-duality is an exact symmetry here, the same labelling scheme carries over to the NS5-F1 frame and its $AdS_3$ limit.

%%%%%%%%%%%%%%%%%%%%%%%%%%%%%%%%%%%%%%%
%%%%%%%%%%%%%%%%%%%%%%%%%%%%%%%%%%%%%%%
%%%%%%%%%%%%%%%%%%%%%%%%%%%%%%%%%%%%%%%
%%%%%%%%%%%%%%%%%%%%%%%%%%%%%%%%%%%%%%%

\section{Discussion}
\label{sec:discussion}

We have seen that the null gauging of WZW models provides a uniform way to construct NS5-brane backgrounds in worldsheet string theory.  A useful feature of this approach is that it yields directly the effective target space geometry.  Perhaps the most intricate examples explored to date are the intersecting fivebrane examples of GMM type analyzed in section~\ref{sec:3d intersection}.  In this concluding section, we discuss some features of the above models, and some avenues for further research.

%%%%%%%%%%%%%%%%%%%%%%%%%%%%%%%%%%%%%%%
\subsection{D-branes and 4d gauge theory}

The GMM model of section~\ref{sec:3d intersection} provides a worldsheet realization of the sorts of intersecting brane constructions which have been used to engineer a variety of 4d gauge theories (see~\cite{Giveon:1998sr} for a review).  A key ingredient of these constructions is the introduction of D-branes stretching between/from the fivebranes and KK monopoles~-- D3-branes in type IIB to generate 2+1 gauge theories, and D4-branes in type IIA to generate 3+1 gauge theories.  It might be interesting to investigate such D-branes in the GMM model and the resulting open string dynamics; aspects of such branes were explored in~\cite{Quella:2002fk,Quella:2003kd,Israel:2005fn,Martinec:2019wzw}.

In the null gauging approach, D-branes in the projected ``downstairs'' geometry begin their life as D-branes in the unprojected ``upstairs'' geometry whose worldvolume is invariant under the isometries being gauged.  The simplest D-branes in the WZW model are associated to conjugacy classes in the group, for instance for $\sutwo$ these are two-spheres.  The brane worldvolume carries a magnetic flux whose quantization restricts the size of the two-sphere to discrete values.  In the construction of~\cite{Martinec:2019wzw}, if not already invariant under the gauge isometry such branes were extended along the gauge orbits by a smearing technique~\cite{Quella:2002fk,Quella:2003kd}.  If the upstairs D-brane worldvolume passes through a locus where the gauge orbits degenerate (in the two-sphere example, the orbits are the azimuthal circles degenerating at the two poles), then in the downstairs geoometry the D-brane appears to ``end'' at that location~-- the projected D-brane consists of just the polar direction, as all points on the azimithal circle are gauge-equivalent.  Because the dilaton in the projected geometry reflects the volume of the gauge orbits, it blows up at the degeneration locus, which is then the (smeared) locus of fivebrane sources (as we see in equation~\eqref{GMM Geff}-\eqref{GMM Phieff}.
One thus has a realization of D-branes ending on NS5-branes.
The quantization of the underlying conjugacy classes translates into the discrete positions of the fivebranes along the smeared source locus (the $\chi$ circle in the GMM model)~-- although supergravity smears out the fivebrane source location, and perturbative string theory can only localize sources to within the string length, D-branes probe the fivebrane locations with Planckian precision.

In the GMM model, one can have D-branes stretching between fivebranes in the same stack which ought to be described by the same sorts of branes constructed in~\cite{Martinec:2019wzw}.   D-branes stretching between NS5-branes in the two different stacks, or between NS5-branes and KK monopoles, are likely to be more complicated, but are of interest due to their connection to the intersecting brane construction of 3d and 4d gauge theories~\cite{Giveon:1998sr}.  

In the Argyres-Douglas models of section~\ref{sec:ArgyresDouglas}, the analogous D-branes are 1/2-BPS states in the effective gauge theory of NS5-branes wrapped on a Seiberg-Witten curve
We leave the detailed exploration of these various D-branes to future work.

%%%%%%%%%%%%%%%%%%%%%%%%%%%%%%%%%%%%%%%
\subsection{The correspondence limit}

An interesting feature of the more complicated fivebrane backgrounds explored here is the possibility to reach and surpass the correspondence limit for strings in $AdS_3$ and linear dilaton backgrounds~\cite{Giveon:2005mi}.  In the correspondence regime one doesn't really trust the effective geometry, for instance in the $AdS_3$ case the correspondence point occurs when $\rads = \lstr$, \ie\ when $k_\sl=1$. 

For the Argyres-Douglas and GMM cosets, one can reach the correspondence limit for the lowest allowed levels of the $\sutwo$ factors, namely $k_\su=2$.  Lower values of the level are disallowed by the requirement of a non-negative level $k_\su\tight-2$ of the underlying bosonic $\sutwo$ current algebra; this then feeds into the $\sltwo$ level $k_\sl$ through the central charge constraint~\eqref{sltwo level} for the Argyres-Douglas models, and~\eqref{n0 value} for the GMM model.  However for other examples \eg\ three stacks of fivebranes intersecting over $\bR^{1,1}$, or the non-critical models of~\cite{Giveon:1999zm} describing fivebranes wrapped a complex surface times $\bR^{1,1}$ (discussed briefly in section~\ref{sec:2d version}), one can exceed the correspondence limit.
In the triple intersection example, the levels are related by 
\be
\frac 1{k_\sl} = \frac 1{k_\su} + \frac 1{k_\su'} +  \frac 1{k_\su''} ~,
\ee
so in particular setting all the $\sutwo$ levels to equal two gives $k_\sl = 2/3$.  For the non-critical models compactifying the fivebrane on an algebraic surface singularity down to $\bR^{1,1}$, the levels are related by $k_\sl^{-1} = 1 +  k_\su^{-1} $ and so one is always below the correspondence point, and again one can get get down to $k_\sl=2/3$ by setting the $\sutwo$ level to its minimum allowed value.

As one approaches the correspondence point, a number of features of the spectrum undergo a transition.  The number of closed string states bound to the cap in the geometry decreases as the range of allowed values~\eqref{jsl range} of $\sltwo$ spin $\jsl$ narrows for discrete series states $\cD_{\jsl}^\pm$. 
The D-brane states also undergo a remarkable change as one crosses the correspondence point, for instance it was shown in~\cite{Kutasov:2005rr} that a probe ``hairpin'' brane suspended from the asymptotic region $\rho\to\infty$ has a wavefunction that delocalizes in the linear dilaton throat as one approaches the correspondence point from above.

Below the correspondence point, the throat is quite stringy -- the slope of the linear dilaton is large in string units, and the angular space is completely stringy.  There are very few local operators bound to the cap, in particular no BPS modes that deform the Coulomb branch vevs and the only shape deformations of the supertube cap are RR modes.  
There is however a continuum of scattering states, which will propagate down the throat and return with a phase shift determined by the two-point function of the ``upstairs'' group $\cG$~\cite{Giveon:2015cma,Martinec:2017ztd,Martinec:2018nco}.  

The situation is qualitatively quite similar to the 1+1d non-critical string (see for instance~\cite{Martinec:2004td} for a review).  There one has a bulk scalar mode (in the bosonic or type 0A string; two in the case of type 0B) scattering off a wall described by $\cN=0$ or $\cN=1$ Liouville theory on the worldsheet.  Spacetime consists of time together with the ``radial'' direction described by the Liouville field.  The slope of the Liouville theory's linear dilaton puts the background on the stringy side of the correspondence transition, which is why there are no black holes in the spectrum.  The models are known to be dual to a matrix quantum mechanics, which in hindsight is the earliest known instance of gauge/gravity duality.  Gauging the angular degrees of freedom of the matrix eliminates all the angular modes, and reduces the dynamics down to the collective dynamics of the eigenvalues only.%
\footnote{Even if one doesn't gauge these modes away, they are sufficiently energetic that they decouple in the infrared scaling limit that defines the model~\cite{Gross:1990md}.}
In other examples of gauge/gravity duality it is these angular modes which are responsible for black hole entropy, and their absence provides a rationale for the absence of black holes in the spectrum.  The eigenvalues of the matrix themselves are realized as D-branes of the dual string theory description.

%%%%%%%%%%%%%%%%%%%%%%%%%%%%%%%%%%%%%%%
\subsection{The Calabi-Yau/Landau-Ginsburg correspondence}

All of the fivebrane backgrounds we have been discussing here exhibit a non-compact version of the Calabi-Yau/Landau-Ginsburg correspondence%
~\cite{Martinec:1988zu,Vafa:1988uu}, 
related to FZZ duality.  In this correspondence, the geometrical background of the supergravity solution sourced by the fivebranes (as well as strings/momenta in the supertube backgrounds) embodied in the sigma model effective action, is inextricably linked to a winding string condensate in the background, embodied in a worldsheet superpotential.  The $\sutwo$ WZW model gives rise to an $\cN=2$ Landau-Ginsburg model, while the $\sltwo$ factor gives rise to $\cN=2$ Liouville theory, with a combined superpotential 
\be
\cW = \sfZ^{k_\su} + \lambda_0 \, e^{k_\su \sfX} ~.
\ee
FZZ duality is the strong/weak coupling duality between the $\sltwo/\uone$ coset model and $\cN\tight=2$ Liouville theory~\cite{Ooguri:1995wj,Giveon:1999px}.  The coset model is weakly coupled at large $k_\sl$ while Liouville theory is weakly coupled at small $k_\sl$; the crossover is the correspondence point $k_\sl=1$.  Typically there is mixing of string modes under worldsheet RG flow, which grows strong as the curvature increases; weakly-coupled intuitions about what is a graviton background and what is a winding tachyon background break down when the curvature is of order string scale, and for $k_\sl<1$ it is more accurate to describe the background as a winding string condensate given by the Liouville potential than as a geometry given by the sigma model metric and $B$-field.  For instance it was argued in~\cite{Giveon:2005mi} that black hole states are not normalizable in the regime $k_\sl<1$, and that the high energy density of states is instead dominated by a Hagedorn gas of fundamental strings.

The string backgrounds below the correspondence transition studied here are best described via an $\cN=2$ Liouville wall, coupled to $\bR^{1,1}$ and a stringy angular space.  The effective theory will live in $\bR^{2,1}$ times a squashed angular space, with a linear dilaton in one of the noncompact spatial directions and an exponential wall keeping perturbative strings out of the strong coupling region.  There will be a finite spectrum of light, field theoretic modes as well as a tower of stringy excitations.  The latter are a new feature as compared to the known matrix models of non-critical string theory, whose 1+1d target spacetime doesn't allow transverse oscillator excitations of the string and so the perturbative string spectrum just consists of a few field-theoretic modes. 
Again there are no black holes in the spectrum.  It is natural to ask whether the D-branes in these backgrounds play a similar role in elucidating their non-perturbative structure; this question is currently under investigation.

Indeed, it was argued in~\cite{Elitzur:2000pq,Israel:2005fn,Martinec:2019wzw} that the D-branes in these models are avatars of the nonabelian little string of Little String Theory, made massive by pulling apart the fivebranes onto their Coulomb branch.  In the black hole regime, the black fivebrane entropy is accounted for by the Hagedorn entropy of little strings~\cite{Maldacena:1996ya}.  As the D-brane realization of little strings on the Coulomb branch shows, the little string excitations are confined to the fivebrane worldvolume.  In the non-critical fivebrane models, the fivebranes wrap a collapsed cycle in a non-compact Calabi-Yau, but the cycle is so tiny that it remains invisible to the strings and D-branes that probe the wrapped fivebranes.  As a consequence, there are fewer than four polarizations of little string oscillations along the fivebrane worldvolume~-- two in the case of the Argyres-Douglas theories, and none in the compactifications down to $\IR^{1,1}$ described briefly in section~\ref{sec:2d version}.  The absence of black holes in the spectrum for these latter models seems to be related to the absence of physical oscillator excitations of the little string, so that there is little to no entropy of the nonabelian degrees of freedom usually responsible for black hole microstates.  Again this structure is reminiscent of the 1+1d non-critical string, where the D-branes describing the non-perturbative physics have no nonabelian excitations, and there are no black holes in the spectrum.

A throughline of the discussions in~\cite{Martinec:2017ztd,Martinec:2018nco,Martinec:2019wzw,Martinec:2020gkv} (see also~\cite{Martinec:2015pfa}) is that the formation of a black hole horizon in the fivebrane system is a Coulomb-Higgs transition in the underlying brane dynamics, where the entropy of nonabelian little string fluctuations becomes available and starts to dominate.  We see that the passage through the correspondence transition is correlated to a suppression of the phase space available to little string fluctuations, such that only the Coulomb branch is available.  

There are thus many parallels between the known matrix model of 2d string theory and noncritical little string theory below the correspondence transition, in parrticular the absence of a ``deconfinement'' transition of nonabelian brane dynamics and therefore the absence of a black hole spectrum, and an effective bulk description in terms of Liouville theory in a stringy regime.  One might hope again for a solvable model for the non-gravitational dual theory.

%%%%%%%%%%%%%%%%%%%%%%%%%%%%%%%%%%%%%%%
\subsection{Comments on symmetric product spacetime CFT's}

If there is a tractable field theory dual to the models discussed here, it is very likely in the regime where the nonabelian little string dynamics is suppressed.  Indeed, the known examples of spacetime CFT's dual to fivebrane backgrounds all lie in a stringy regime of this sort, and have the form of a symmetric product orbifold.

Consider for instance a single stack of fivebranes on $\cM\times\bS^1$, where $\cM=\bT^4$ or $K3$; there is a corner of the spacetime CFT moduli space which has a description as a nonlinear sigma model on the symmetric product orbifold $(\cM)^N/S_N$.  
The effective dual description of this corner of the moduli space involves a background with $N$ fundamental strings and a {\it single} fivebrane~\cite{Larsen:1999uk} (and hence a regime where there is no nonabelian fivebrane physics).

The structure of this moduli space was analyzed in~\cite{Seiberg:1999xz,Larsen:1999uk}, where a rather explicit parametrization of the moduli space of the usual $AdS_3\times\bS_3\times\bT^4$  spacetime CFT was given.  The idea is that the U-duality group of this background is the subgroup $\Gamma$ of $O(5,5;\bZ)$ that preserves the background onebrane-fivebrane charge vector.  This subgroup is strictly smaller than what one might naively have thought to be the little group of a charge vector, namely $O(5,4;\bZ)$ (setting aside the case where $N$ is prime, so that by construction $n_5\tight=1,n_1\tight=N$).  On the other hand, the moduli space is the symmetric space $O(5,4)/O(5)\tight\times O(4)$, modulo the U-duality group $\Gamma$.  Since $O(5,4;\bZ)$ is larger than $\Gamma$ but acts effectively on the moduli space, elements of $O(5,4;\bZ)$ not in $\Gamma$ map a point in the moduli space to some other point in the moduli space, and at the same time don't preserve the charge vector of the background.  The element does however preserve the symplectic inner product on charges, which is $N$.  Thus every unordered pair $n_1,n_5$ such that $N=n_1n_5$ characterizes a cusp of the moduli space, and there are elements of $O(5,4;\bZ)$ that map any given cusp to another one somewhere else in the moduli space with a different unordered pair.  One can argue that the symmetric product corresponds to the pair $n_1=N, n_5=1$ by analyzing the spectrum of wrapped branes and momenta on $\bT^4$.  But one can move continuously through the moduli space to one of the other cusps where the effective description has $n_5$ large, and work in the regime of moduli where the effective description has NS fluxes.  Then the radius of curvature of $AdS_3$ and $\bS^3$ are of order $\sqrt{n_5}$ in string units, and the high energy density of states is dual to an effective geometry with a horizon.  But in the regime described by the symmetric orbifold, one is at the correspondence point and the notions of ``geometry'' and ``horizon'' are inappropriate due to the stringiness of the background.

Similarly, the large $\cN=4$ theory on $AdS_3\times\bS^3\times\bS^3\times\bS^1$ is a decoupling limit of two stacks of fivebranes intersecting in codimension eight with fundamental strings bound to the intersection.  It was proposed in~\cite{Argurio:2000tb} that the relevant spacetime CFT has a symmetric orbifold locus in its moduli space based on the symmetric product of the $\sutwo\times\uone$ WZW model.  This suggestion was further analyzed in~\cite{Gukov:2004ym,Eberhardt:2017pty,Eberhardt:2019niq}.
Here, however, the U-duality group is almost trivial; there are only two moduli~-- the string coupling and the RR axion~-- and the only identifications of the moduli space are the period shifts of the RR axion, so there is only a single cusp of the moduli space.  The data of the symmetric product once again matches the background where one of the fivebrane charges is equal to one; the charge on the other stack is the level of the $\sutwo$ WZW model in the symmetric product.  Because the U-duality group is so limited, there is not a regime in the moduli space where the $AdS_3$ curvature scale can be made large relative to the string scale, and one is always below the correspondence point~-- the relation of $\sltwo$ and $\sutwo$ levels
\be
\frac 1{k_\sl} = \frac 1{k_\su} + \frac 1{k_\sutil}
\ee
leads to $k_\sl<1$ when we set one of the $\sutwo$ levels equal to one.  

It is not hard to see why any symmetric product describes a stringy background~\cite{Banks:1998dd}.  Consider its spectrum in the regime of energies $1\ll L_0 \ll c/24$.  The most entropic way to spend this energy budget is to take half of the energy $E$ to create the longest possible cyclic twist of the tensor factors, and then to take the remaining energy and spend it on oscillator excitations; the effective central charge of the long cyclic twisted sector is proportional to the energy, and the density of states in this sector is well into the Cardy regime, thus
\be
S \sim 4\pi\sqrt{(c_{\rm eff}/12)E} \propto E ~~.
\ee
Starting from very low energy (of order the AdS scale), the microcanonical ensemble in this regime has a Hagedorn density of states, and so one must be in a regime where the dynamics is quite stringy.

This simple analysis is consistent with the known examples.  For the single stack of fivebranes compactified on $\bT^4$ or $K3$, the curvature radius of $AdS_3$ at the symmetric product locus in moduli space is precisely the string scale because the background has only a single fivebrane, and thus one is at the correspondence point; for the large $\cN=4$ models, when one of the fivebrane fluxes is equal to one, the $AdS_3$ radius is less than one in string units.

The non-critical models based on a null gauging of $\sltwo\times\sutwo\times\bR^{1,1}$ also lead to a stringy throat, and a suppression of nonabelian little string dynamics.  Indeed it was proposed in~\cite{Giveon:1999zm} (see equation 4.3 of that paper, and also appendix A of~\cite{Giveon:2005mi}) that adding fundamental strings to the background of these models of fivebranes compactified on an fourfold version of the conifold would lead to string theory on $AdS_3\times\bS^1$ times an $\cN=2$ minimal model.  The continuum of wound strings on the Coulomb branch has a Fock space structure paralleling that of a symmetric product orbfold.  Our results here flesh out some aspects of this proposal.
The supertube gauging leads to an $AdS_3$ cap on the linear dilaton fivebrane throat, and describes particular 1/2 BPS spacetime backgrounds.  
Of course one should not trust the geometry here at all, since all these backgrounds are below the correspondence transition and better described by Liouville theory than by geometry.  But it is perhaps this very feature that encourages the search for a symmetric orbifold realization of the dual spacetime CFT.

Symmetric products of a variety of other SCFT's have been studied in~\cite{Argurio:2000tb,Gaberdiel:2019wjw,Eberhardt:2019qcl,Belin:2020nmp}, particular those with $\cN=2$ supersymmetry.   Again one may ask what could be a candidate gravity dual.  We expect a bulk $AdS_3\times \bS^1$ whose isometries realize the bosonic sector of global $\cN=2$ superconformal symmetry, times the target space $\cM$ in the symmetric product.  
For instance,~\cite{Belin:2020nmp} consider $\cM$ to be an $\cN=2$ minimal model, which can be realized as the WZW coset model $\frac{\sutwo}{\uone}$.
It would be interesting to explore further the possibilities of determining the spacetime CFT dual to some of the models discussed here.

%%%%%%%%%%%%%%%%%%%%%%%%%%%%%%%%%%%%%

%%%%%%%%%%%%%

\vspace{3mm}

\section*{Acknowledgements}
%%%%%%%%%%%%%

We thank 
Bruno Balthazar,
Amit Giveon,
David Kutasov,
Stefano Massi,
and
David Turton
for useful discussions.
The work of EJM and TDB is supported in part by DOE grant DE-SC0009924.
TDB is also supported by a Mafalda and Reinhardt Oehme Fellowship of the Enrico Fermi Institute at the University of Chicago.

\vspace{6mm}

\appendix

%%%%%%%%%%%%%%%%%%%%%%%%%%%%%%%%%%%%
%%%%%%%%%%%%%%%%%%%%%%%%%%%%%%%%%%%%

%\newpage
\section{Conventions}
\label{app:conventions}

%%%%%%%%%%%%%%%%%%%%%%%%%%%%%%%%%%%%

In this appendix we record several group theory and current algebra conventions.
\vspace{-1mm}

\refstepcounter{subsection}
\subsection*{\thesubsection \quad SU(2)} \label{sec:sutwoapp}
%\subsection{SU(2)}\label{sec:sutwoapp}

We parametrize SU(2) via Euler angles as
\be \label{eq:su2-param}
g_{\su} ~=~ e^{\frac{i}{2}(\psi-\phi)\sigma_3}e^{i \theta \sigma_1}e^{\frac{i}{2}(\psi + \phi)\sigma_3} 
~= ~
\begin{pmatrix} \cos\theta \;\! e^{i\psi}  &~~  i \:\!  \sin\theta\:\!  e^{-i\phi} \\ 
i \:\!  \sin\theta\:\!  e^{i\phi}  &~~ \cos\theta \;\! e^{-i\psi} \end{pmatrix}\,
\,.
\ee
These are the conventions used in~\cite{Martinec:2018nco} (note that~\cite{Martinec:2017ztd} had conventions related to these by $\phi \to - \phi$).
Here the $\sigma_a$ are the usual Pauli matrices, explicitly
\be
\sigma_1 = \left(\begin{array}{cc} 0 &1 \\ 1 &
    0 \end{array}\right), \qquad \sigma_2 = \left(\begin{array}{cc} 0 &-i \\ i &
    ~~0  \end{array}\right), \qquad \sigma_3 = \left(\begin{array}{cc} 1 &~~0 \\ 0 &
    -1  \end{array}\right) \,.
\ee
The generators and structure constants of the Lie algebra $\su(2)$ are as usual
\be
T^a_\su = \frac{1}{2}\sigma_a \,, \qquad \quad \T{(f_\su)}{ab}{\:\! c} = i\:\! \epsilon_{abc} \,, \qquad  \epsilon_{123}=1\,.
\ee

The supersymmetric $\sutwo$ level $\nfive$ current algebra consists of current operators $ J_\su^a$ and their fermionic superpartners $\psi_\su^a$ having the OPE structure
\begin{align}
J_\su^a(z)\, J_\su^b(0) &\sim \frac{\frac12 \nfive \, \delta^{ab}}{z^2} + \frac{i\epsilon_{abc} J_\su^c(0)}{z}
\nn\\
J_\su^a(z)\, \psi_\su^b(0) &\sim i\epsilon_{abc}\frac{\psi^c_\su(0)}{z}
\\
\psi_\su^a(z)\, \psi_\su^b(0) &\sim \frac{\delta^{ab}}{z}
\nn
\end{align}
with the Killing metric $\delta^{ab} = {\rm diag}(+1,+1,+1)$.  One can define a set of ``bosonic'' $\sutwo$ level $\nfivetil=\nfive-2$ currents $j_\su^a$ that commute with the fermions,
\be
j_\su^a = J_\su^a + \frac i2\epsilon_{abc}\psi_\su^b \psi_\su^c  ~.
\ee
The primary fields $\Psihat_{\jsu \msu \bmsu}$ of the current algebra have conformal dimensions
\be
h = \bar h = \frac{\jsu(\jsu+1)}{\nfive} ~;
\ee
unitarity restricts the allowed spins $\jsu$ of the underlying bosonic current algebra to the range
\be
\label{su2 reps app}
\jsu = 0,\frac12,\dots,\frac{\nfive}{2} -1 ~.
\ee

The primary operators of the supersymmetric theory are built by combining primaries $\Psi_{\jsu \msu \bmsu}$ of the level $\nfivetil$ bosonic current algebra with level two primaries of the fermions (namely the identity operator $\One$, spin operator $\Sigma$, and the fermions themselves).  The highest weight fields are of three types:
\be
\label{su2hwr}
\Psihat_{jjj} = \Psi_{jjj}
~~,~~~~
\Psihat_{j+\half,j+\half,j+\half} = \Psi_{jjj}\Sigma_{++}
~~,~~~~
\Psihat_{j+1,j+1,j+1} = \Psi_{jjj}\psi^+\bar\psi^+
~~,
\ee
with the remaining operators of the zero mode multiplet obtained through the action of the zero modes of the total current $J^-_\su$.  In building massless string states, one uses the purely bosonic highest weight operator for NS sector states whose polarization does not lie along $\sutwo$, and the highest weight operator with a fermion attached when the polarization does lie along $\sutwo$; Ramond sector operators involve $\Sigma$, with the various spinor polarizations reached through the action of the zero modes of $\psi^-_\su,\bar\psi^-_\su$.  In the type II string, the choice of fermion decoration is independent on left and right, and there is a chiral GSO projection onto odd total fermion number in the matter sector.

These operators have a {\it parafermion} decomposition%
\footnote{Our notation here largely follows~\cite{Martinec:2001cf}, see also~\cite{Giveon:2015raa}, except that we work in conventions where $\alpha'=1$, so that T-duality is $R\to 1/R$, instead of the convention $\alpha'=2$ of those works.}
under the current $J_\su^3$%
~\cite{Fateev:1985mm,Gepner:1986hr,Gepner:1987qi}
obtained by extracting the dependence on $J^3_\su, \bar J^3_\su$.  To this end, one bosonizes the currents
\begin{align}
j^3_\su=i\sqrt{\nfivetil}\,\partial Y_\su 
~&,~~~~
\psi^+_\su \psi^-_\su = i\sqrt2\,\partial H_\su ~,
\nn\\
J^3_\su=i\sqrt{\nfive}\,\partial \cY _\su
~&,~~~~
J_\cR^\su = \frac\nfivetil\nfive \psi^+_\su\psi^-_\su - \frac2\nfive j^3_\su
= i\sqrt{\frac{2\nfivetil}{\nfive}} \, \partial \cH_\su ~,
\end{align}
and similarly for the right-movers.  The current $J^3_\su$ forms a $U(1)$ supermultiplet with the fermion $\psi^3_\su$, and every operator in the super-WZW model can be written as a product of a parafermion operator and an operator from the super-$U(1)$ theory.

The relation between the canonically normalized bosons $H,Y$ and $\cH,\cY$ for $\sutwo$ is a rotation in field space
%given for instance in~\cite{Martinec:2001cf} (see also~\cite{Chang:2014jta}):
\begin{align}
Y &= \sqrt{\frac{\nfivetil}\nfive}\,\cY - \sqrt{\frac{2}{\nfive}} \,\cH
\nn\\
H &= \sqrt{\frac{\nfivetil}\nfive}\,\cH - \sqrt{\frac{2}{\nfive}} \,\cY  ~.
\end{align}
The parafermion decomposition of $\sutwo$ primaries $\Psi$, parafermions $\Lambda$, and their super-versions $\medhat\Psi$ and $\medhat\Lambda$ is then given by
\begin{align}
\Psi_{jm}^{w} &= \Lambda_{jm} \, \exp\Bigl[i\frac{2}{\sqrt{\nfivetil}} \Bigl(m+\frac{\nfivetil}2 w\Bigr) Y_\su\Bigr]
\nn\\[.3cm]
\medhat \Lambda_{jm}^{\tilde\eta} &= \Lambda_{jm}^{~} \exp\Bigl[ i\sqrt{\frac{8}{\nfive\nfivetil}} \,\Bigr(-m+\frac{\nfivetil}{2}\tilde\eta\Bigr)\cH_\su\Bigr]
\nn\\[.4cm]
\medhat\Psi_{jm}^{\eta.w} &= \Psi_{jm}^{w} \exp\Bigl[i \sqrt2\,\eta H\Bigr] =  
\medhat \Lambda_{jm}^{\tilde\eta} \exp\Bigl[ i\frac{2}{\sqrt\nfive} \Bigl(m+\tilde\eta+\frac\nfive2 w\Bigr)\cY_\su \Bigr] ~.
\end{align}
where $\tilde\eta=\eta-w$.  Spectral flow relates $\sutwo$ representations as follows:
\be
\medhat \Lambda_{j,j}^{\eta-w=0}=\medhat \Lambda_{\frac\nfivetil2-j , j-\frac\nfivetil2}^{\eta-w=-1}
~~,~~~~
\medhat \Psi_{j,j}^{\eta=0,w=0}=\medhat \Psi_{\frac\nfivetil2-j , j -\frac\nfivetil2}^{\eta=0,w=+1}
\ee
where $\nfivetil=\nfive-2$.

%%%%%%%%%%%%%%%%%%%%%%%%%%%%%%%%%%%%%%%

\refstepcounter{subsection}
\subsection*{\thesubsection \quad SL(2)} \label{sec:sltwoapp}
%\subsection{SL(2)}\label{sec:sltwoapp}

We parametrize $\sltwo$ as $SU(1,1)$ via
\be
\label{eq:sl2-param}
g_{\sl} \;=\; e^{\frac{i}{2}(\tau-\sigma)\sigma_3}e^{\rho \sigma_1}e^{\frac{i}{2}(\tau + \sigma)\sigma_3} \,.
\ee
Again these are the same conventions used in~\cite{Martinec:2018nco} (related to those of~\cite{Martinec:2017ztd} by $\sigma \to - \sigma$).\\
The generators and structure constants of the Lie algebra $\su(1,1)$ are 
\be
T^1_{\sl} \,=\, \frac{i}{2}\sigma_1 \,, \quad~ T^2_{\sl} \,=\, \frac{i}{2}\sigma_2 \,, \quad~ T^3_{\sl} \,=\, \frac{1}{2}\sigma_3 \;; \qquad~~%
 \T{(f_{\sl})}{12}{3} \,=\, -i \,, \quad \T{(f_{\sl})}{23}{1} \,=\, \T{(f_{\sl})}{31}{2} \,=\,  i \,.
\ee

The supersymmetric $\sltwo$ level $\nfive$ current algebra consists of currents $J_\sl^a$ and their fermionic superpartners $\psi_\sl^a$ having the OPE structure
\begin{align}
J_\sl^a(z)\,J_\sl^b(0) &\sim \frac{\frac12 \nfive \,h^{ab}}{z^2} +  \frac{\T{(f_{\sl})}{ab}{c} J_\sl^c(0)}{z}
\nn\\
J_\sl^a(z)\, \psi_\sl^b(0) &\sim \T{(f_{\sl})}{ab}{c}\frac{\psi^c_\sl(0)}{z}
\\
\psi_\sl^a(z)\, \psi_\sl^b(0) &\sim \frac{h^{ab}}{z}
\nn
\end{align}
with the Killing metric $h^{ab} = {\rm diag}(+1,+1,-1)$.  One can similarly define a set of ``bosonic'' $\sltwo$ level $\nfivehat\!=\!\nfive\!+\!2$ currents $j_\sl^a$ that commute with the fermions,
\be
j_\sl^a = J_\sl^a + \frac 12 \T{(f_{\sl})}{a}{bc}\psi_\sl^b \psi_\sl^c  ~.
\ee
The primary fields $\Phihat_{\jsl \msl \bmsl}$ of the current algebra have conformal dimensions
\be
h = \bar h = -\frac{\jsl (\jsl-1)}{\nfive} ~.
\ee

These operators also have a superparafermion decomposition under the current $J_\sl^3$%
~\cite{Dixon:1989cg,Griffin:1990fg,Dijkgraaf:1991ba}%
\footnote{Again our notation here largely follows~\cite{Martinec:2001cf}, see also~\cite{Giveon:2015raa}.}
obtained by extracting the dependence on $J^3_\sl, \bar J^3_\sl$.  To this end, one bosonizes the currents
\begin{align}
j^3_\sl=-i\sqrt{\nfivehat}\,\partial Y_\sl 
~&,~~~~
\psi^+_\sl \psi^-_\sl = i\sqrt2\,\partial H_\sl ~,
\nn\\
J^3_\sl=-i\sqrt{\nfive}\,\partial \cY _\sl
~&,~~~~
J_\cR^\sl = \frac\nfivehat\nfive \psi^+_\sl\psi^-_\sl + \frac2\nfive j^3_\sl
= i\sqrt{\frac{2\nfivehat}{\nfive}} \, \partial \cH_\sl ~,
\end{align}
and similarly for the right-movers.  
Note that the bosons $\cY,\bar\cY$ and $Y,\bar Y$ are timelike.  The bosons $H,Y$ and $\cH,\cY$ are then related by a boost in field space: 
%given for instance in~\cite{Martinec:2001cf} (see also~\cite{Chang:2014jta}):
\begin{align}
Y &= \sqrt{\frac{\nfivehat}\nfive}\,\cY + \sqrt{\frac{2}{\nfive}} \,\cH
\nn\\
H &= \sqrt{\frac{\nfivehat}\nfive}\,\cH + \sqrt{\frac{2}{\nfive}} \,\cY  ~.
\end{align}

The parafermion decomposition of $\sltwo$ primaries $\Phi$, parafermions $V$, and their super-versions $\medhat\Phi$ and $\medhat V$ is then given by
\begin{align}
\Phi_{jm}^{w} &= V_{jm} \, \exp\Bigl[i\frac{2}{\sqrt{\nfivehat}} \Bigl(m+\frac{\nfivehat}2 w\Bigr) Y_\sl\Bigr]
\nn\\[.3cm]
\medhat V_{jm}^{\hat\eta} &= V_{jm}^{~}\, \exp\Bigl[ i\sqrt{\frac{8}{\nfive\nfivehat}} \,\Bigr(m+\frac{\nfivehat}{2}\hat\eta\Bigr)\cH_\sl\Bigr]
\nn\\[.4cm]
\medhat\Phi_{jm}^{\eta.w} &= \Phi_{jm}^{w}\, \exp\Bigl[i \sqrt2\,\eta H\Bigr] =  
\medhat V_{jm}^{\hat\eta} \,\exp\Bigl[ i\frac{2}{\sqrt\nfive} \Bigl(m+\hat\eta+\frac\nfive2 w\Bigr)\cY_\sl \Bigr] ~.
\end{align}
where $\hat\eta=\eta+w$.

The conformal dimension of the $\sltwo$ primary $\Phihat_{\jsl \msl \bmsl}^{\sl}$ decomposes as
\begin{equation}
\label{slpfspec-app}
h\big[\medhat V_{\jsl \msl \bmsl}\big] = \frac{-\jsl (\jsl-1)+\msl^2}{\nfive} 
~,~~~~
\bar h\big[\medhat V_{\jsl \msl \bmsl}\big] = \frac{-\jsl (\jsl-1)+\bmsl^2}{\nfive} ~,
\end{equation}
with the rest made up by the dimension of the $\cY,\bar\cY$
exponentials.  Again the fields $V_{\jsl \msl \bmsl}$ commute with the current $J^3_\sl$, and so are the natural building blocks for representations of the gauged theory.
The shift of the $J^3_\sl$ charge $\msl \to (\msl \!+\!\frac12 \nfive \wsl)$ leads to the flowed conformal dimension 
\be
h\bigl[V^{(\wsl,\bwsl)}_{\jsl \msl \bmsl}\bigr] = -\frac{\jsl
  (\jsl-1)}{\nfive} - \msl \wsl - \frac{\nfive}{4}  \wsl^2 \;.
\ee
Because we are working on the universal cover of $\sltwo$, the $\tau$ direction is non-compact and so there is no independent left and right spectral flow, but rather a simultaneous spectral flow that shifts $\msl,\bmsl$ by the same amount.

Unitary representations of bosonic $\sltwo$ current algebra are as follows. One has the principal discrete series (on both left and right)
\begin{equation}
\cD_j^+ = \bigl\{ \ket{j,m}~\bigl| ~  j\in\IR_+\, ;~~ m\!=\! j+n\, ,~~n\in\IN \bigr\}
\end{equation}
and its conjugate 
\begin{equation}
\cD_j^- = \bigl\{ \ket{j,m}~\bigl| ~  j\in\IR_+\, ;~~ m\!=\! -(j+ n)\,
,~~ n\in\IN \bigr\} \,,
\end{equation}
restricted to the range
\be
\frac 12 < j < \frac{\nfive+1}{2} ~;
\end{equation}
in addition one has the continuous series representations $\cC_j^\alpha$ (again on both left and right)
\begin{equation} \label{eq:cont-series-app}
\cC_j^\alpha = \bigl\{ \ket{j,m}~\bigl| ~  j\!=\!
\coeff12(1+i\nu)\, ,~~\nu \in\IR\, ;~~m\!=\! \alpha+n\, ,~~n\in\IZ\, ,~~0\!\le\alpha\!<1\in\IR \bigr\} ~.
\end{equation}

%%%%%%%%%%%%%%%%%%%%%%%%%%%%%%%%%%%%
%%%%%%%%%%%%%%%%%%%%%%%%%%%%%%%%%%%%

%%%%%%%%%%%%%%%%%%%%%%%%%%%%%%%%%%%%%%%%%%%%%%%%
%%%%%%%%%%%%%%%%%%%%%%%%%%%%%%%%%%%%%%%%%%%%%%%%

%%%%%%%%%%%%%%%%%%%%
\section{Spacetime supersymmetry in gauged WZW models}
\label{app:spacetimesusy}

In the critical dimension examples studied in%
~\cite{Martinec:2017ztd,Martinec:2018nco,Martinec:2019wzw,Martinec:2020gkv}, 
the worldsheet operators that implement spacetime supersymmetry can be cast in terms of the representations of the $O(10,2)$ current algebra formed by the worldsheet fermions.  This current algebra has an equivalent bosonic realization obtained through the standard bosonization relations
\begin{align}
\psi^+_\sl\psi^-_\sl = i\sqrt2\partial H_\sl
~~,~~
\psi^+_\su\psi^-_\su &= i\sqrt2\partial H_\su
~~,~~
\psi_3^\sl\psi_3^\su = i\sqrt2\partial H_3
\nn\\
\psi_t\psi_y = i\sqrt2\partial H_{ty}
~~,~~~~
\chi^{~}_6\chi^{~}_7 &= i\sqrt2\partial H_{67}
~~,~~~~
\chi^{~}_8\chi^{~}_9 = i\sqrt2\partial H_{89}
\end{align}
where $\chi^{~}_i$ are the worldsheet superpartners of the $\bT^4$ coordinates $X^i$, $i=6,7,8,9$.  The operators
\be
\label{O(10,2) spinfield}
S_\varepsilon = \exp\Big[ \frac{i}{\sqrt2}\Big(\varepsilon_\sl H_\sl + \varepsilon_\su H_\su + \varepsilon_3 H_3 + \varepsilon_{ty} H_{ty} + \varepsilon_{67} H_{67} + \varepsilon_{89} H_{89} \Big) \Big] ~,
\ee
where each $\varepsilon_a=\pm1$ are the Dynkin indices of the spinor polarization, transform in the spinor representation of $O(10,2)$.  One can impose the Majorana-Weyl condition $\prod_a \varepsilon_a=\pm1$ independently on left- and right-movers to describe either type IIA or type IIB.  For the present discussion we will assume $\prod_a \varepsilon_a=+1$.
The spacetime supersymmetry operators in the $-\half$ picture are then%
\footnote{As discussed in~\cite{Giveon:1998ns}, there are two canonical constructions of supercharges in $AdS_3\times\bS^3$~-- the standard spin field of the orthogonal group current algebra of the fermions, dressed by worldsheet ghosts; and the spin field generated by spectral flow in a worldsheet $\cN=2$ $\cR$-symmetry~\cite{Banks:1987cy}.  In the case of Coulomb branch NS5-brane backgrounds, these two constructions differ by a null gauge transformation.  The supersymmetry currents built from~\eqref{O(10,2) spinfield} have suitable OPE's with all vertex operators, while the supersymmetry current built using $\cN=2$ spectral flow may have $\bZ_{\nfive}$ branch singularities with some vertex operators outside the physical spectrum.
}
\be
\label{susy charge}
Q_\varepsilon^{ (-\half)} = \oint e^{-(\varphi-\tilde\varphi)/2} S_\varepsilon ~,
\ee
where we have bosonized the superghosts $\beta,\gamma$ and spinor ghosts for null gauging $\tilde\beta,\tilde\gamma$ according to%
\footnote{Note that we use here a bosonization prescription slightly different from the original formulation laid out in~\cite{Friedan:1985ge}; this improved version~\cite{Martinec:1988bg,Takama:1988it,Horowitz:1988xf,Horowitz:1988ip} has the advantage that it  (a) is manifestly $\cN=2$ supersymmetric, (b) has a manifestly $\cN=2$ bosonization, and (c) is BRST invariant in both the ``large'' and ``small'' Hilbert spaces of the bosonic ghosts.  
%In this formalism, the different pictures arise from attaching delta-function localized supermoduli to a given vertex operator.
}
\begin{align}
\beta\gamma= -\partial\varphi
~~,~~~~
\gamma &=  e^\varphi \partial\xi
~~,~~~~
\beta = e^{-\varphi} \eta
\nn\\
\tilde\beta\tilde\gamma = -\partial\tilde\varphi
~~,~~~~
\tilde\gamma &=  e^{\tilde\varphi} \partial\tilde\xi
~~,~~~~
\tilde\beta = e^{-\tilde\varphi} \tilde\eta
\end{align}
in order to construct the ghost spin field $e^{-(\varphi-\tilde\varphi)/2}$ that intertwines NS and R sector ground states for the ghosts.  
There will thus be ``pictures'' of physical state vertex operators for both the $\beta,\gamma$ and $\tilde\beta,\tilde\gamma$ systems.%

The BRST charge 
\begin{align}
\label{BRST charges}
\cQ = \oint \big( cT \tight+ \gamma G \tight+ \tilde c\cJ \tight+ \tilde\gamma \lamb + \textit{ghosts} \big)
\end{align}
imposes several constraints on the supersymmetry charge~\eqref{susy charge}.  In the supercurrent term $\gamma G$, the $f_{abc}\psi^a\psi^b\psi^c$ terms in the $\sltwo$ and $\sutwo$ WZW models impose the constraints%
\footnote{Our conventions agree with those of~\cite{Itzhaki:2005tu} Appendix B.}
\be
\label{chirality constraints}
\varepsilon_\sl\,\varepsilon_\su\,\varepsilon_3 = -1
~~,~~~~
\varepsilon_{ty}\,\varepsilon_{67}\,\varepsilon_{89} = -1  
\ee
where we have fixed an overall spinor parity in $O(10,2)$.
The $\tilde c\cJ$ term imposes the constraint
\be
l_1\,\varepsilon_\sl + l_2\,\varepsilon_\su = 0 ~.
\ee
Finally, the fermionic null constraint $\tilde\gamma\lamb$ involves
\begin{align}
\label{supernull current}
\lamb &= 
l_1 \psi^3_\sl+ l_2\psi^3_\su + l_3\psi^t +l_4 \psi^y 
\\[.2cm]
&= 
\frac{e^{i\sqrt{2} H_3}}{\sqrt{2}}\big(l_1+l_2\big) + \frac{e^{-i\sqrt{2} H_3}}{\sqrt{2}}\big(-l_1+l_2\big) +
\frac{e^{i\sqrt{2} H_{ty}}}{\sqrt{2}}\big(l_3+l_4\big) + \frac{e^{-i\sqrt{2} H_{ty}}}{\sqrt{2}}\big(-l_3+l_4\big) \,.
\nn
\end{align}
we set $l_1=l_2$ and $l_3=-l_4$, with the latter vanishing for the Coulomb branch but nonzero for supertubes.  This constraint imposes the restriction that the spinor polarization be transverse to the null vector; acting on the supercharge~\eqref{susy charge}, the result is that one has $\varepsilon_3=+1$ always, and for supertubes additionally one requires $\varepsilon_{ty}=-1$.

Altogether, one finds that the supercharge $Q_\beta$ lies in the BRST cohomology for
\be
\label{spinor charges}
\big(q_\varphi,q_{\tilde\varphi};\varepsilon_\sl,\varepsilon_\su,\varepsilon_3 \big) =  \big( -\hf,+\hf;\varepsilon,-\varepsilon,-1 \big) 
~~,~~~~
\varepsilon_{ty}\, \varepsilon_{67}\, \varepsilon_{89} = -1 
~~,~~~~
\varepsilon =\pm1 ~,
\ee
with in addition $\varepsilon_{ty}=-1$ for supertubes.
Together with the analogous right-moving counterparts one has (8,8) supersymmetries for the Coulomb branch and (4,4) spacetime supersymmetry for the supertube, as expected.
As usual, picture changing~\cite{Friedan:1985ge} yields versions of these supersymmetry charges in other superselection sectors of the ghost numbers measured by the currents $\beta\gamma$ and $\tilde\beta\tilde\gamma$.
The end result is a perturbative string S-matrix that satisfies the constraints of spacetime supersymmetry.

%
%%%%%%%%%%%%%%%%%%%%%%%%%%%%%%%

Spacetime supersymmetry in the worldsheet formalism can also be engineered from $\cN\tight=2$ worldsheet supersymmetry~\cite{Gepner:1987qi,Banks:1987cy}, and this construction is useful for the more general backgrounds considered here.
Gauged WZW models provide a broad class of examples of worldsheet $\Neqtwo$ theories~\cite{Gepner:1987qi,Kazama:1988qp}.%
\footnote{The construction of spacetime supercharges in such theories was considered in%
~\cite{Giveon:1999jg,Berenstein:1999gj,Giveon:2003ku} in the context of string solutions involving $AdS_3$ with NS fluxes.  As discussed in~\cite{Giveon:1998ns}, there are two canonical constructions of spacetime supercharges, one associated to the Killing spinors of AdS3 and another using the worldsheet $\Neqtwo$ which breaks the $AdS_3$ symmetry.  Since the null gauging breaks that symmetry through the choice of a null Killing vector, we are here not interested in spacetime supersymmetry associated to $AdS_3$ isometries; instead, we will be interested in the worldsheet $\Neqtwo$ construction.}
 
Let $\cM$ be a worldsheet theory with a $\uone$ current $J_0$, such that $\coset=\cM/\uone$ has $\Neqtwo$ worldsheet supersymmetry, with $\cR$-symmetry current $J_\cR$; we bosonize these currents as
\be
J_\cR = ia_\coset\,\partial\cH_\coset
~~,~~~~
J_0 = i\sqrt{1-a_\coset^2}\,\partial\cY_\coset
\ee
where
\be
a_\coset^2 = \frac13 c_\coset
\ee
In addition we bosonize the fermionic partners of the $\sltwo$ and $J_0$ currents
\be
\psi_\sl^+\psi_\sl^- = i\sqrt 2 \,\partial H_\sl
~~,~~~~
\psi_\sl^3\psi^0 = i\sqrt 2 \,\partial H_3
\ee
and note that the $\sltwo/\uone$ parafermion theory has an $\Neqtwo$ supersymmetry with $\cR$-symmetry current
\be
J_\cR^\sl = i a_\sl\,\partial\cH_\sl
\ee
where $a_\sl^2 = \frac13 c_\sl = 1\tight+\frac2{k_\sl }$.
Finally, the two $\uone$ factors $J_3^\sl, J_0$ together with their superpartners $\psi_3,\psi_0$ form a free $\Neqtwo$ system with $c\tight=3, a\tight=1$.

When the string worldsheet theory admits global $\Neqtwo$ superconformal symmetry (only the $\cN\tight=1$ of which is gauged, of course), there is a canonical construction of spacetime spin fields using spectral flow in the $\Neqtwo$ $\cR$-symmetry.
One can  make spin fields of the form~\cite{Gepner:1987qi,Banks:1987cy}
\begin{align}
\label{STsusy}
\hat S_\vareps &= \exp\Big[ \frac{i}{\sqrt2}\Big(\vareps_\sl \, a_\sl \cH_\sl + \vareps_3 H_3 + \vareps_\coset a_\coset \cH_\coset\Big) \Big]
\end{align}
where $\vareps=\pm1$, by spectral flow of the identity operator using the $\cN\tight=2$ $\cR$-symmetry.  

We can relate the spin field $S_\vareps$ of~\eqref{O(10,2) spinfield} to this $\Neqtwo$ spin field as follows, in a way that generalizes to all the models considered in this paper (we write explicit formulae for the models of section~\ref{sec:ArgyresDouglas}, but the examples of sections~\ref{sec:1d-2d intersection} and~\ref{sec:3d intersection} are straighforward as well).
One has $\matter = \sutwo\times \bR_t^{~}\times \bS^1_y\times \bT^4$, and each factor contributes to $\cH_\coset$ in proportion to its central charge
\begin{align}
a_\coset \cH_\coset &= a_\su\cH_\su + \hat\vareps_{ty}H_{ty} + \vec\vareps_\IT\cdot \vec H_{\IT}
\nn\\[.2cm]
\cY_\coset &= \cY_\su
~~,~~~~
\psi_0=\psi^3_\su
~~,~~~~
\psi_t\psi_y = i\sqrt2\,\partial H_{ty} 
\end{align}
where $a_\su^2 = \coeff13 c_\su = 1\tight- \frac{2}{k_\su}$, with $k_\su$ the level of the supersymmetric $\sutwo$ current algebra; and $\vec H_\IT$ bosonize the fermions on an internal torus compactification of the fivebrane worldvolume~-- $\IT^4$ for the critical dimension, $\IT^2$ for the Argyres-Douglas models, and nothing for the 2d noncritical models. 
For the definitions of $\cH_\su,\cY_\su$ see appendix~\ref{app:conventions}. 
In the critical theory, comparing to~\eqref{O(10,2) spinfield} and using the relations
 \be
H_\sl = a_\sl \cH_\sl + \sqrt{a^2_\sl-1}\,\cY_\sl  
~~,~~~~
H_\su = -a_\su \cH_\su + \sqrt{1-a^2_\su}\,\cY_\su  
\ee
to compare the $\cH$ charges,
one finds $\varepsilon_\su=-\varepsilon_\coset$, $\varepsilon_{ty}=\vareps_\coset \hat\vareps_{ty}$ 
and similarly for the spin polarizations on the four-torus $\IT$; for the noncritical theories one again has solutions to the BRST constraints analogous to~\eqref{spinor charges}, with $\varepsilon_{ty}\, \varepsilon_{67}\, \varepsilon_{89} = -1 $ replaced by the appropriate chirality constraint on the dimensionally reduced fivebrane worldvolume.

In the critical dimension, $S_\varepsilon$ differs from $\hat S_\varepsilon$ by an exponential of $\cY_\sl,\cY_\su$ in the proportion of the bosonized null current
\be
\label{cY for spin fields}
V_\cY = \exp\Big[ \frac{i}{\sqrt{2}}\sqrt{a_\sl^2-1}\Big(\vareps_\sl \cY_\sl + \vareps_\cK\cY_\cK \Big)\Big] 
\ee
because one has $a_\sl^2-1 = 1-a_\su^2$ as a consequence of the fact that $k_\sl=k_\su$.  
When one has $k_\sl\ne k_\su$, as in the constructions considered here, then the contributions of $\cH$ and $\cY$ do not obviously combine so neatly; but because of the squashing of $\sutwo$, the exponentials of $\cY_\su$ and $\cY_\sl$ continue to come in the null combination~\eqref{cY for spin fields} that appears in the chiral-GSO projected theory.  In terms of the parametrization~\eqref{pLR su squashed}, this exponential has $p_L$ with $n=0,\ell=\pm1$ and $p_R=0$.  As usual, this operator will have branch cuts with respect to the operators in the original spectrum from section~\ref{sec:ADgeom} which are removed by the chiral GSO projection, and we then add the NS-R and R-NS spectrum generated from the remaining operators by the supercharges 
\eq{
Q = \oint \hat S_\varepsilon V_\cY ~.
}

%%%%%%%%%%%%%%%%%%%%%%%%%%%%%%%%%%%%%%%%%%%%%%%%%%%%%%
%%%%%%%%%%%%%%%%%%%%%%%%%%%%%%%%%%%%%%%%%%%%%%%%%%%%%%

\section{Non-null gauging alternative}
\label{sec:non-null gauging}

In this appendix, we point out that there is an alternative non-null choice of gauged isometries that leads to the same spectrum, at least in the Argyres-Douglas and GMM models and probably more broadly.  Consider the gauging of $\sltwo\times\sutwo$ at levels $k_\sl,k_\su$, with the choice of gauge currents
\eq{
\cJ_1 = J^3_\sl - \frac{k_\sl}{k_\su} J^3_\su 
~~&,~~~~
\bar\cJ_2 = -\bar J^3_\sl + \bar J^3_\su
\\[.1cm]
\bar\cJ_1 = \bar J^3_\sl - \frac{k_\sl}{k_\su} \bar J^3_\su
~~&,~~~~
\cJ_2 = +J^3_\sl - J^3_\su 
}
from which one reads off the corresponding Killing vectors $\xi^a_{L,R}$, $a=1,2$.
Note that in the critical dimension, $k_\sl=k_\su$ and these currents are null; otherwise they are not null, and the gauging procedure is a different route to the same physics.  Note also that neither group is squashed.  The gauge group is again rank two, but now none of the components of the gauge fields decouple.
Following the procedure outlined in section~\ref{sec:nullreview}, the term quadratic in gauge fields is
\eq{
&\half A_1\bar A_1\big(\xi^1_L\tight+\xi^1_R\big)\tight\cdot \big(\xi^1_L\tight+\xi^1_R\big)
+ A_1\bar A_2\, \xi^1_L \tight\cdot \xi^2_R
+ A_2\bar A_1\, \xi^2_L \tight\cdot \xi^1_L 
+ \half A_2\bar A_2\big(\xi^2_L\tight+\xi^2_R\big)\tight\cdot \big(\xi^2_L\tight+\xi^2_R\big)
\\[.1cm]
&\hskip 1cm
=2\frac{k_\sl}{k_\su} \, A_1\bar A_1\big( k_\sl\cos^2\theta - k_\su \cosh^2\rho \big)
+ 2 A_2\bar A_2 \big( k_\su\sin^2\theta + k_\sl \sinh^2\rho \big)
\\[-.1cm]
&\hskip 3cm
+ 2k_\sl\big(A_1\bar A_2 - A_2\bar A_1\big)\, \big( \cos^2\theta - \cosh^2\rho \big) ~~.
}
Upon integrating out the gauge fields, one arrives at the same effective geometry~\eqref{AD effective geom}-\eqref{AD Phieff}, and the gauge constraints yield the same spectrum of BPS operators~\eqref{AD chiralops}.  Indeed, the \naive\ singularity in the effective geometry at $\rho=\theta=0$ is already apparent in the degeneration of the above quadratic form in gauge fields.  One can find a similar set of non-null currents for the GMM model that again yield the same results as null gauging.

This choice of gauge group is perhaps a somewhat more precise way of formulating the model.  As noted in~\cite{Martinec:2018nco}, the timelike gauge orbits are generally non-compact, and so the gauge group is typically $\bR$ for the timelike vector gauging of currents like $\cJ_1,\bar\cJ_1$, and a compact $\uone$ for the spacelike axial gauging of currents like $\cJ_2,\bar\cJ_2$.

%%%%%%%%%%%%%%%%%%%%%%%%%%%%%%%%%%%%%%%%%%%%%%%%%%%%%%
%%%%%%%%%%%%%%%%%%%%%%%%%%%%%%%%%%%%%%%%%%%%%%%%%%%%%%
%%%%%%%%%%%%%%%%%%%%%%%%%%%%%%%%%%%%%%%%%%%%%%%%%%%%%%
%%%%%%%%%%%%%%%%%%%%%%%%%%%%%%%%%%%%%%%%%%%%%%%%%%%%%%

%%%%%%%%%%%%%%%%%%%%%%%%%%%%%%%%%%%%
%%%%%%%%%%%%%%%%%%%%%%%%%%%%%%%%%%%%

\newpage
\vskip 1cm

\bibliographystyle{JHEP}      

\bibliography{microstates}

\end{document}